\documentclass[11pt]{article}
\usepackage{latexsym}

\newcommand{\be}{\begin{equation}}
\newcommand{\ee}{\end{equation}}
\newcommand{\bea}{\begin{eqnarray}}
\newcommand{\eea}{\end{eqnarray}}
\newcommand{\nn}{\nonumber\\}

\newcommand{\etal}{{\it et al.\/}}
\newcommand{\lhs}{{\it l.h.s.\/}} 
\newcommand{\rhs}{{\it r.h.s.\/}} 
\newcommand{\ie}{{\it i.e.\/}}

\renewcommand{\d}{\partial}
\newcommand{\dif}{{\rm d}}
\newcommand{\la}{\langle}
\newcommand{\ra}{\rangle}
\newcommand{\CFT}{{\mbox{\tiny \rm CFT}}}
\newcommand{\eff}{{\mbox{\tiny \rm eff}}}

\newcommand{\D}{{\cal D}}
\renewcommand{\O}{{\cal O}}
\newcommand{\eps}{\epsilon}
\newcommand{\gind}{{g_{\mbox{\tiny \rm ind}}}}
\newcommand{\prop}{{\cal K}}

\makeatletter 
\def\secteqno{\@addtoreset{equation}{section}% 
\def\theequation{\thesection.\arabic{equation}}}

\begin{document}

\secteqno
\baselineskip=.58cm

%%%%%%%%%%%%%%%%   TITLE    %%%%%%%%%%%%%%%%%%%%
%
\renewcommand{\thefootnote}{\fnsymbol{footnote}}

{\hfill \parbox{4cm}{ 
        MIT-CTP-3133 \\ 
        hep-th/0105048 }} 

\bigskip\bigskip

\begin{center} \LARGE \bf Randall-Sundrum models and the regularized
AdS/CFT correspondence
\end{center}

\bigskip\bigskip

\centerline{
Manuel P\'{e}rez-Victoria\footnote[1]{manolo@lns.mit.edu} }

\bigskip
\bigskip
\centerline{ \it Center for Theoretical Physics}
\centerline{ \it Massachusetts Institute of Technology}
\centerline{ \it Cambridge, {\rm MA}  02139}
\bigskip
\bigskip
\bigskip

\renewcommand{\thefootnote}{\arabic{footnote}}

\centerline{\bf Abstract} 
\begin{center}
\parbox{11.5cm}{It has been proposed that Randall-Sundrum models can be
holographically described by a regularized (broken) conformal field
theory. We analyze the foundations of this duality using a regularized
version of the AdS/CFT correspondence. We compare two- and three-point
correlation functions and find the same behaviour in both
descriptions. In particular, we show that the regularization of the 
deformed CFT generates kinetic terms for the sources, which hence
can be naturally treated as dynamical fields. We also discuss the
counterterms required for two- and three-point correlators in the
renormalized AdS/CFT correspondence.}  
\end{center}

\newpage

%%%%%%%%%%%%% TEXT %%%%%%%%%%%%%%%%%%%

\section{Introduction}

The work of L.~Randall and
R.~Sundrum has brought about
a great interest in models with warped extra dimensions. 
With appropriate stabilization
mechanisms~\cite{Goldberger:1999uk,DeWolfe:2000cp}, some 
of these models can account for the large hierarchy between the
Planck and the electroweak scales~\cite{Randall:1999ee}. On the other
hand, noncompact extra dimensions may be compatible with current
observations when the space-time is warped~\cite{Randall:1999vf}.

In the Randall-Sundrum models and their generalizations, space-time is
a piece of AdS. It is natural, hence, to try to take advantage of the
AdS/CFT
correspondence~\cite{Maldacena:1998re,Gubser:1998bc,Witten:1998qj},
which relates a quantum field theory  
in $d$ dimensions to a theory in $d+1$ dimensions including gravity. 
J.~Maldacena has suggested~\cite{Maldaunpub} that the noncompact
Randall-Sundrum model (RS2) is dual to 4-D gravity coupled to a
strongly coupled conformal theory with an ultraviolet
cutoff. This idea has been realized more explicitly in
\cite{Gubser:2001vj,Giddings:2000mu,Giddings:2000ay}  
(see also~\cite{Verlinde:2000fy,Verlindesug} for a string-theoretical
construction),  
where it is further argued that 4-D gravity arises from the local
counterterms that are required to define finite AdS amplitudes in the
AdS/CFT correspondence. Other developements in this direction
include~\cite{Wittensug,Duff:2000mt,deHaro:2000wj}.  
This holographic description has been invoked in a number of works
to study different aspects of Randall-Sundrum
models~[17--34]. % This is a citation
%  \cite{Garriga:2000bq,Das:2001zc}. 
More recently,
the duality has been extended to the Randall-Sundrum compact model
(RS1) in~\cite{Arkani-Hamed:2000ds,Rattazzi:2001hs}. This has allowed
the authors of these papers 
to discuss important phenomenological aspects of RS1 using the
holographic description.

The purpose of this paper is to study in greater detail the
basis of this duality. To start with, we shall argue that the content 
of the duality itself must be qualified: the regularization of the CFT
generates kinetic terms for the 4-D gravity. This provides an
attractive holographic picture of Randall-Sundrum models, since
propagating gravity arises from regularization, just as in the AdS
description. We shall support this interpretation
with evidence from the calculation of correlation functions in both
sides of the correspondence. Specifically, we shall be
interested in correlation functions in the effective theory where
all the bulk degrees of freedom have been integrated out (in a
classical approximation).
In order to compare with the CFT, it will be convenient to pose the gravity
calculations in a way that resemble the standard AdS/CFT
calculations. Our results are also relevant to the renormalized
AdS/CFT
correspondence~\cite{Henningson:1998gx,Balasubramanian:1999re,
deHaro:2001xn,Chalmers:2000gc} and we shall make some observations
about the renormalization of the CFT in this context. 

The layout of the paper is as follows. In Section~\ref{regRS}
we argue that RS2 is dual to a regularized CFT which automatically
contains gravity. In Section~\ref{AdS} we develop the effective
formalism for a bulk scalar field in RS2 and calculate explicitly two
and three-point functions. We discuss the relation to the Kaluza-Klein
description and perform an expansion in the position of the Planck
brane, which allows comparison with the CFT calculations. In
Section~\ref{CFT} we describe the regularization of
the CFT in the AdS/CFT correspondence, and compare with the AdS
calculations of the previous section. We also discuss what
counterterms are required to renormalize the CFT. In Section~\ref{RS1}
we study the duality for RS1. In this case we only calculate explicitly
two-points functions. In Section~\ref{localized} we study sources
localized on the Planck brane or inside the bulk. We finish with some
general comments in Section~\ref{discussion}.

%%%%%%%%%%%%%%%%%%%%%%%%%%%%%%%%%%%%%%%%%%%%%%%%%%%%%%%%

\section{Randall-Sundrum as a regularized CFT}
\label{regRS}

There are only a few examples of the AdS/CFT correspondence in which
both the gravity and the boundary theories are known. The best
stablished one is the equivalence of ${\cal N}=4$ super Yang-Mills in
4 flat dimensions and 
type IIB string theory on $AdS_5\times S_5$, especially in the limit
of strong 't~Hooft coupling and large N, in which the gravity theory
reduces to classical type IIB supergravity. Other interesting cases 
with less number of supersymmetries and even nonconformal have also
been studied. More generally, it is commonly believed that any
consistent gravity theory on an asymptotically AdS space has some kind
of conformal dual. Here we assume that this is the case, at least for
the gravity theories in Randall-Sundrum models. 

We shall work in Euclidean $AdS_{d+1}$ and use Poincar\'e coordinates,
which in the Euclidean case cover the whole space. The metric
reads 
\be
\dif s^2 = \frac{L^2}{z_0^2} (\dif \vec{z}^2 + \dif z_0^2) \, ,
\ee
where $\vec{z}$ is a vector with components $z_i,~i=1,\ldots,d$,
$z_0 \geq 0$ and $L$ is the AdS curvature. The dual field theory can
be thought of as living on the boundary of this space, which
is a $d$ dimensional sphere located at $z_0=0$ (together with the
point at infinity). 
In the usual AdS/CFT correspondence, the
holographic theory does not 
contain dynamical gravity. Essentially, the reason is that
normalizable graviton modes in the bulk cannot reach the boundary of
$AdS$, due to the divergent behaviour of the metric as $z_0\rightarrow
0$. Moreover, the metric does not induce a determined metric at the
boundary, but a conformal class. From the point of view of the CFT,
the graviton is an external source coupling linearly to the
stress-energy tensor. It does not propagate.

In RS2 a $d$ dimensional brane is located at $z_0=\eps$ for some
$\eps\geq 0$ and space-time is restricted to $z_0\geq
\eps$. Actually, in the standard RS2 one glues another copy
of this semi-infinite AdS space and imposes a $Z_2$ symmetry, with
$z_0=\eps$ a fixed point of the orbifold, but for simplicity we shall
work with just one copy of the amputated AdS, with orbifold boundary
conditions at $z_0=\eps$. Our results then differ by a factor 1/2 from
the ones in the complete orbifold (for fields that are even under the
$Z_2$). Note that the actual value of $\eps$ is not relevant, since we
can always redefine the coordinates $\vec{z}$ to set 
$\eps=L$. However, we shall be interested in an expansion in
$\eps$, so it is convenient to keep it independent of
$L$. An important feature of this construction is that a well-defined
metric is induced on this brane (usually called Planck brane) and
that normalizable graviton modes can reach it. The same holds for other
bulk fields. Therefore, the holographic dual of RS2 is expected to
contain $d$-dimensional dynamical gravity and, in general, propagating
degrees of freedom corresponding to the different bulk fields.
On the other hand, the Planck brane abruptly terminates the space
beyond $z_0=\eps$. The ultraviolet/infrared (UV/IR)
correspondence~\cite{Susskind:1998dq} 
indicates that this corresponds to some kind of cutoff in the
holographic theory. Let us make these ideas more explicit.

The dynamical content of the AdS/CFT correspondence is given by the
identity \cite{Gubser:1998bc,Witten:1998qj}
\be
W^\CFT[\varphi] = S^\eff[\varphi]  \, , \label{AdSCFTrelation}
\ee
where
\bea
W^\CFT[\varphi] & = & - \ln \int \D A \, \exp \{- S^\CFT [A] + \varphi
\O \} \, , \\  \label{WCFT}
S^\eff[\varphi] & = & - \ln \int_{\phi(0)=\varphi} \D \phi \,
\exp \{-S[\phi] \} \, .
\eea
Here, $\phi$ represents any of the fields propagating in AdS (in
particular, the graviton), $\O$ is the corresponding dual operator in
the CFT, $\varphi$ is a field on the boundary manifold which
acts as a source for $\O$, and $A$ stands for all the
fields in the CFT. $S^\CFT$ and $W^\CFT$ are,
respectively, the action and the generating functional of connected
correlation functions in the CFT, whereas $S$ is the action of
the gravity theory. The corresponding effective action, $S^\eff$, is a
functional of $\varphi$ defined by a path integral---we use field
theory notation---in which the fields are constrained to take definite
values at the boundary.  We are interested in situations in which
classical gravity 
is a good description of RS2, which corresponds to strong 't~Hooft
coupling and large N on the CFT side (if the CFT is a gauge
theory, as in the known examples of the correspondence). In this case,
$S^\eff$ reduces to the classical on-shell action, subject to the
constraint that the boundary values of the fields $\phi$ coincide with
$\varphi$. 

As a matter of fact, both sides of the relation~(\ref{AdSCFTrelation})
are ill-defined. The on-shell gravity action is divergent in the
IR whereas the correlation functions obtained from $W^\CFT$
contain UV divergencies at coincident points (again, we
see the UV/IR correspondence at work). In order to make sense of this
identity we introduce an IR regulator in the AdS theory and a
UV regulator in the CFT. In particular, we can introduce
an IR cutoff in AdS restricting the space to the
range $z_0 \geq \eps$, just as in RS2. This corresponds to some
definite (unknown) UV regularization on the CFT side. 
Then, the regularized AdS/CFT correspondence states that 
\be
W_\eps^\CFT[\varphi] = S_\eps^\eff[\varphi]  \, ,
\label{regAdSCFTrelation} 
\ee
with
\bea
W_\eps^\CFT[\varphi] & = & - \ln \int_\eps \D A \, \exp \{- S^\CFT [A]
+ \varphi \O \} \, , \\
S_\eps^\eff[\varphi] & = & - \ln \int_{\phi(\eps)=\varphi} \D \phi \,
\exp \{-S_\eps[\phi] \} \, , \label{Seff}
\eea
where $S_\eps[\phi]$ is the action of the gravity theory with the space-time
integral restricted to $z_0\geq \eps$, and the $\eps$ in the path
integral indicates regularization. In practice we shall regularize the
correlation functions obtained from functional differentiation of the
unregularized generating functional. The regularized generating
functional is then defined by its expansion in regularized correlation
functions. This modification of the correspondence is relevant for
Randall-Sundrum models because the effective 
theory of RS2 on the Planck brane is nothing but (two copies of) the
\rhs\ of (\ref{regAdSCFTrelation}). Therefore, RS2 is holographically
described by a {\em regularized} CFT. 

Since we have just given heuristic arguments to obtain
(\ref{regAdSCFTrelation}) from (\ref{AdSCFTrelation}), one may have
some doubts about the validity of this proposal. Furthermore, the
effective theory of RS2 is known to contain dynamical gravity and it
is not clear how kinetic terms for $\phi$ appear in $W_\eps^\CFT$. In
order to solve this apparent problem and to better justify the identity
(\ref{regAdSCFTrelation}), we follow
\cite{Gubser:2001vj,Giddings:2000mu,Giddings:2000ay} and
consider the so-called renormalized AdS/CFT correspondence 
\cite{Henningson:1998gx,Balasubramanian:1999re,deHaro:2001xn,Chalmers:2000gc}.
As a by-product we will find the alternative (standard)
interpretation of the holographic theory as a renormalized CFT
coupled to dynamical gravity. The renormalized AdS/CFT correspondence 
reads  
\be
W_{\mbox{\rm \tiny ren}}^\CFT[\varphi] = S_{\mbox{\rm \tiny
ren}}^\eff[\varphi]  \, . \label{renAdSCFTrelation} 
\ee
The \rhs\ is defined by 
\be
S_{\mbox{\rm \tiny ren}}^\eff[\varphi] =
\lim_{\eps\rightarrow 0} \left(S_\eps^\eff[\varphi] + S_\eps^{\mbox{\rm
\tiny ct}} [\varphi] \right) \, , \label{renAdS}
\ee
where $S_\eps^{\mbox{\rm \tiny ct}}$ contains the counterterms required
for a well-defined $\eps\rightarrow 0$ limit. It has been shown
in~\cite{Henningson:1998gx,Balasubramanian:1999re,deHaro:2001xn} that
these counterterms can be written as
functionals of the boundary fields. On the other hand, a
renormalized CFT is generically defined by the addition of
local UV counterterms:
\bea
W_{\mbox{\rm \tiny ren}}^\CFT[\varphi] & = &
-\lim_{\eps^\prime\rightarrow 0} \ln \int_{\eps^\prime} \D A \, \exp
\left\{- S^\CFT [A] + \varphi \O -S^{\CFT \,\mbox{\rm \tiny
ct}}_{\eps^\prime} [A,\varphi] \right\} \nn
& = & \lim_{\eps^\prime\rightarrow 0}
\left( W_{\eps^\prime}^\CFT[\varphi] + W_{\eps^\prime}^{\CFT
\,\mbox{\rm \tiny ct}}[\varphi] \right) \, , \label{renCFT}
\eea
where the prime in $\eps^\prime$ indicates that this regularization
may be unrelated to the one in the AdS theory. $W_{\eps^\prime}^{\CFT
\,\mbox{\rm \tiny ct}}$ can be nonlocal. 
Eqs.~(\ref{renAdS}) and (\ref{renCFT}) define valid renormalization
procedures, and it has been checked that
(\ref{renAdSCFTrelation}) holds in some examples (in particular, the
same Weyl anomaly is obtained in both theories
\cite{Henningson:1998gx}). Note that both 
sides of the relation are scheme dependent, since the finite
part of the counterterms is not fixed. Now, we can go one step back,
remove the limits and state that
\bea
S_\eps^\eff[\varphi] & = &  W_{\eps^\prime}^\CFT[\varphi] +
W_{\eps^\prime}^{\CFT \,\mbox{\rm 
\tiny ct}}[\varphi]-S_\eps^{\mbox{\rm \tiny ct}}[\varphi]  \nn
&\equiv& W_\eps^\CFT[\varphi] \label{basic}
\, .
\eea
The first identity is a consequence of (\ref{renAdSCFTrelation}), up
to terms that vanish when $\eps\rightarrow 0$ and in the second one we
have simply defined a new regularization of the CFT by adding specific
counterterms to the generating functional. We see
that a regularization scheme for the CFT exists such that
(\ref{regAdSCFTrelation}) is fulfilled, at least to order $\eps^0$. 
To this order, (\ref{basic}) can also be written as
\be
S_\eps^\eff[\varphi] =  W_{\mbox{\rm \tiny ren}}^\CFT[\varphi] -
S_\eps^{\mbox{\rm \tiny ct}}[\varphi] \, . \label{basicren}
\ee
Although $S_\eps^{\mbox{\rm \tiny ct}}$ can be nonlocal, we shall see
that the \rhs\ of (\ref{basicren}) can be written as a path integral
weighted by the exponential of minus a local action
$S^\CFT[A]+S^\prime[A,\varphi]$. This gives an alternative holographic
interpretation of RS2 as a {\em 
renormalized} CFT coupled to the theory $S^\prime$. Note that
$S_\eps^{\mbox{\rm \tiny ct}}$, and then $S_\eps^\prime$, contain in
particular the $d$-dimensional Einstein-Hilbert action.
Of course, the very same $-S_\eps^{\mbox{\rm \tiny ct}}$ arises in the
$\eps$ regularization of the CFT. Hence, an appropriately regularized CFT
automatically contains dynamical $d$-dimensional gravity and, more generally,
dynamical $d$-dimensional fields corresponding to all bulk fields that
do not vanish on the brane. In the next sections we shall check
explicitly that this is indeed the case. One should keep in mind
that (\ref{basicren}) is only a good approximation for small enough
$\eps$, which as we shall see corresponds to small
momenta compared with the inverse AdS curvature. However, there is a
double counting in thinking of RS2 as a cutoff CFT coupled to
dynamical gravity.  

Since all these arguments are independent of any modification of the theory
at large values of $z_0$, the same conclusions apply to
RS1. Of course, in this case the second brane breaks conformal
invariance in the IR.

%%%%%%%%%%%%%%%%%%%%%%%%%%%%%%%%%%%%%%%%%%%%%%%%%%%%%%%%%%%%%%

\section{Correlations functions in RS2}
\label{AdS}

In this section we calculate correlation functions in
RS2 for the fields induced on the Planck brane, using perturbation
theory in the coupling constants. For 
simplicity we consider only scalar fields but analogous results
hold for fields with higher spin. Moreover, we assume that these
scalars move in a fixed background, \ie, we ignore the back-reaction
on the metric of the scalars. In our perturbative
formalism, this would be described as higher-point correlators of
scalars and gravitons in linearized gravity.

\subsection{General formalism}
\label{genform}

To simplify the notation, we
start with a single scalar field, $\phi$. Let $\varphi(\vec{z}) =
\phi(\eps,\vec{z})$. We are interested in the effective theory for the
boundary field $\varphi$. The correlation functions of several fields
$\phi$ can be obtained from the generating functional
\bea
Z[j] & = & \int \D \phi \, \exp \left\{ - S[\phi] + \int_{\mbox{\rm
\tiny brane}} \varphi j \right\} \nn
& = & \int \D \varphi \exp \left\{ - S^\eff[\varphi] + \int_{\mbox{\rm \tiny brane}}
\varphi j \right\} \, ,  \label{splitting}
\eea
where $S^\eff[\varphi]$ is defined as in (\ref{Seff}). From now on the
subindex $\eps$ is implicit. We have splitted the path integral into two
parts: first, we calculate the effective action by integrating out the
bulk degrees of freedom with Dirichlet boundary conditions; second, we
use it to calculate $Z[j]$. $S^\eff$ contains nonlocal 
interactions. For classical induced fields, it coincides with
the generating functional of 1PI correlation functions. Therefore, 
\be
\la \varphi(x_1) \cdots \varphi(x_n) \ra^{\rm 1PI} = \left[ \frac{\delta}{\delta
\varphi(x_1)} \cdots \frac{\delta}{\delta \varphi(x_n)} \,
S^\eff[\varphi] \right]_{\varphi=0} \, ,
\ee
and the relation (\ref{regAdSCFTrelation}) implies
\be
\la \varphi(x_1) \cdots \varphi(x_n) \ra^{\rm 1PI} = \la \O(x_1)
\cdots \O(x_n) \ra \, . \label{1PIconnected}
\ee
The \rhs\ is a connected correlation function in the dual
CFT. The connected correlations functions of boundary fields are
obtained in the standard way from the 1P1 ones. Observe that
(\ref{1PIconnected}) means that, as expected for dual fields,
the generating functionals of $\varphi$ and $\O$ correlation functions
are related by a Legendre transformation.

At this point, one might worry about the relation between the
correlators calculated from $S^\eff$, which are obtained with
Dirichlet boundary conditions, and the propagators calculated in
\cite{Garriga:2000yh,Giddings:2000mu}, which used Neunman boundary
conditions. The analysis below is an extension of the discussion
in~\cite{Giddings:2000mu}. In Randall-Sundrum models one must impose Neumann
boundary conditions at the orbifold fixed points for fields that are
even under the $Z_2$ symmetry. This is implied by the
orbifold structure and continuity of the derivative of the fields at
the branes\footnote{In some important  
cases the action contains terms proportional to delta functions at the
branes that make the field derivatives discontinuous, and one 
must add appropriate corrections. We comment on this below. On the
other hand, odd fields must vanish 
at the Planck brane and do not appear in the effective description}. 

As it was argued in \cite{Giddings:2000mu}, both approaches arise from
two different ways of splitting the path integral. In (\ref{splitting})
the bulk fields are integrated with Dirichlet boundary conditions,
so that one must still perform the integration over the brane
fields. We shall refer to this procedure as the ``D approach''. One could
also integrate first the fields in a neighbourhood of 
the brane, which imposes the constraint that fields should obey Neumann
boundary conditions at the brane. The remaining path integral over the
bulk fields must be carried out with this constraint. We call this
the ``N approach''. Let us consider the classical limit. In the N
approach, the on-shell field generated by a source $j$ located on the
brane can be written as 
\be
\phi(z_0,\vec{z})= \int \dif^{d} x \sqrt{g(\eps)}
\Delta_N(\eps,\vec{x};z_0,\vec{z}) 
j(\vec{x}) \, , \label{Napproach} 
\ee
where $g(\eps)$ is the determinant of the $d+1$-dimensional
metric evaluated on the brane and $\Delta_N$ 
is the scalar Neumann propagator of
\cite{Garriga:2000yh,Giddings:2000mu}. This  
is the Green function of the quadratic equation of motion with a
(bulk) source $J(y)$, with 
boundary conditions $[\d_0 \Delta_N(y_0,\vec{y};
z_0,\vec{z})]_{z_0=\eps}=0$, $y_0\not \! =\eps$,
and $\lim_{z_0\rightarrow \infty} \Delta_N
(y_0,\vec{y};z_0,\vec{z})=0$. We use the notation $\d_0=\d/\d z_0$.
The propagator $\Delta_N(\eps,\vec{x};z_0,\vec{z})$ is defined from
the limit $y_0\rightarrow \eps$; we note that it does not obey the
Neumann condition on the brane. We use in
(\ref{Napproach}) the full metric, instead of the induced metric, in
order to have the same normalization for the propagator as in
\cite{Giddings:2000mu}. Sources inside the bulk will be
considered in Section~\ref{localized}.

In the D approach, the on-shell field is written as
\be
\phi(z_0,\vec{z})= \int \dif^{d} x \prop(\vec{x};z_0,\vec{z})
\varphi(\vec{x}) \, , \label{Dint}
\ee
where $\prop$ is the Dirichlet ``bulk-to-boundary'' propagator, \ie,
a solution to the homogeneous equation of motion obeying
$\prop(\vec{x};\eps,\vec{z})=\delta^{(d)}(\vec{x}-\vec{z})$ and
$\lim_{z_0\rightarrow \infty} \prop=0$. Observe that the field
$\varphi(\vec{x})$ is still off shell. As is customary in AdS/CFT we
have defined $\prop$ such that the unit metric appears in the integral
(\ref{Dint}).  
Both approaches must be
equivalent after complete path integration. Therefore, both
expressions for $\phi$ must agree for on-shell $\varphi$, that is to
say, when the $\varphi$ equation of motion, derived from $S^\eff$, is
imposed. Indeed, in the next subsection we show explicitly that the
propagator of the theory $S^\eff$ is precisely
$\Delta_N(\eps,\vec{x};\eps,\vec{y})$. Hence, on shell,
\be
\varphi(\vec{x})=\int \dif^{d} y \sqrt{g(\eps)}
\Delta_N(\eps,\vec{x};\eps,\vec{y}) j(y) 
\ee
and
\bea
\phi(z_0,\vec{z}) & = & \int \int \dif^d x \dif^d y \, \sqrt{g(\eps)}
\prop(\vec{x};z_0,\vec{z}) \Delta_N(\eps,\vec{x};\eps,\vec{y}) j(y)
\nn 
& = & \int \dif^{d} y \sqrt{g(\eps)} \Delta_N(\eps,\vec{y};z_0,\vec{z})
j(\vec{y}) \, , 
\eea
which agrees with (\ref{Napproach}). We have used the identity 
\be
\int\dif^d x
\Delta_N(\eps,\vec{x};\eps,\vec{y}) \prop(\vec{x};z_0,\vec{z}) =
\Delta_N(\eps,\vec{y};z_0,\vec{z}) \, , \label{glue}
\ee
which can be checked by explicit computation. 
So we see that both procedures give the same bulk fields when the
brane fields satisfy their classical equations of motion. 
As we mentioned in the last footnote, in some cases the action
contains quadratic terms proportional to delta functions (the
so-called boundary mass terms)~\cite{Gherghetta:2000qt}. In the N
approach one has to 
modify the Neunman boundary condition to take into account the
discontinuity of the field derivatives. In the D approach, the
effective action will contain these boundary mass terms, which
change the on-shell $\varphi$ in such a way that the same bulk field
is obtained again. In the following we shall be interested in the
value of $S^\eff[\varphi]$ for off-shell $\varphi$, from which the
correlation functions of these fields can be obtained.
Since correlation functions of $\varphi$ are nothing but correlation
functions of $\phi$ for points on the brane, a direct consequence of
this analysis is that correlation functions calculated in the N
approach are related to (regularized) Witten diagrams by a Legendre
transformation. As in usual field theory, 1PI correlators are easier to
calculate and contain all the information of the theory. Here we are
simply comparing two methods for calculating the same object in the
AdS theory, with no reference to any CFT. But of course,
we are interested in the D procedure because it allows a direct
comparison with the holographic dual.

%%%%%%%%%%%

\subsection{Two-point functions}

Consider a scalar field $\phi$ of mass $M^2$ in $AdS_{d+1}$. The quadratic
part of the Euclidean action reads
\be
S= \frac{1}{2} \int_\eps \dif^{d+1} z  \sqrt{g}
\left(\d^\mu \phi \d_\mu \phi + M^2 \phi^2 \right) \, . \label{quadaction}
\ee
where $\int_\eps \dif^{d+1} z = \int \dif^{d} z \int_\eps^\infty \dif
z_0$. The normalization of $\phi$ has been fixed to obtain a canonical
kinetic term. The equation of motion reads
\be
z_0^{d+1} \d_0 \left( z_0^{-d+1} \d_0 \phi(z_0,\vec{k})\right) - (k^2
z_0^2 + m^2) \phi(z_0,\vec{k}) = 0 \, ,
\ee
where we have Fourier transformed the field in the coordinates tangent
to the brane:
\be
\phi(z_0,\vec{k}) = \int \dif^d z \, e^{i \vec{k} \cdot
\vec{z}} \phi(z_0,\vec{z}) \, .
\ee
Unless otherwise indicated, scalar products are defined with the flat 
Euclidean metric, which is related to the induced metric by a constant
Weyl transformation: $\gind_{ij}=L^2/\eps^2 \, \delta_{ij}$.
Translation invariance along the brane implies that the momentum
$\vec{k}$ is conserved. We have also defined the dimensionless mass
$m^2=L^2 M^2$. This squared mass can be negative, but must obey the
Breitenlohner-Freedman bound $m^2 \geq -d^2/4$. The conformal dimension
of the field (which determines its boundary behaviour) is $\Delta=1/2
( d + \sqrt{d^2+4m^2}) \equiv \nu + d/2$.\footnote{For $-d^2/4 < m^2
\leq -d^2/4 + 1$ there are two 
alternative AdS-invariant quantizations, one with $\Delta$ as defined
above and the other with $\Delta_-=1/2 ( d - \sqrt{d^2+4m^2})$. In this
paper we always consider the first possibility. We shall make some
comments about the second one below.} The particular 
solutions of this differential equation are 
$z_0^{d/2} I_\nu (z_0 k)$ and $z_0^{d/2} K_\nu (z_0 k)$,
where $I_\nu$ and $K_\nu$ are modified Bessel functions and
$k=|\vec{k}|$. Regularity in the interior selects $\phi \propto
z_0^{d/2} K_\nu(z_0 k)$. The Dirichlet bulk-to-boundary
propagator is a solution such that
\be
\lim_{z_0\rightarrow \eps} \prop(z_0,\vec{k}) = 1 \, ; \hspace{1cm}
\lim_{z_0\rightarrow \infty} \prop(z_0,\vec{k}) = 0 \, .
\ee
The explicit form of this propagator is
\be
\prop(z_0,\vec{k})=\left(\frac{z_0}{\eps}\right)^{\frac{d}{2}} \,
\frac{K_\nu(k z_0)}{K_\nu(k\eps)} \, . \label{Dprop}
\ee
The on-shell bulk field is then $\Phi(z_0,\vec{k})=\prop(z_0,\vec{k})
\varphi(\vec{k})$. Inserting this expression in the action,
integrating by parts and using the equation of motion, the (on-shell)
action reduces to a surface term (see~\cite{Freedman:1999tz}, for
instance). Double functional differentiation with respect to $\varphi$
yields 
\be
\la \varphi(\vec{k}) \varphi(\vec{k^\prime}) \ra^{\rm 1PI}=
\delta(\vec{k}+\vec{k^\prime}) L^{-1} \left(\frac{L}{\eps}\right)^d \,
\left(\nu-\frac{d}{2} + \frac{k \eps K_{\nu-1} (k \eps)}{K_\nu(k
\eps)} \right) \, . \label{exacttwopoint}
\ee
From now on we absorb the momentum conservation deltas into the
definition of the correlation functions. The connected two-point
correlator (the propagator of the effective theory) is simply the
inverse of the 1PI one: 
\be
\la \varphi(\vec{k}) \varphi(\vec{-k}) \ra = L
\left(\frac{\eps}{L}\right)^d \, \frac{K_\nu(k
\eps)}{(\nu-\frac{d}{2}) K_\nu(k\eps) + k\eps K_{\nu-1}(k\eps)} \,
. \label{prop}
\ee

It is instructive to study this propagator in Minkowski space. We use
$(-1,1,\ldots,1)$ signature. Taking
provisionally $\eps=L$ and performing a Wick rotation,
the propagator (\ref{prop}) reads
\be
\la \varphi(\vec{k}) \varphi(\vec{-k}) \ra = - L \, \frac{H^{(1)}_\nu(q
L)}{(\nu-\frac{d}{2}) H^{(1)}_\nu(q L) - q L
H^{(1)}_{\nu-1}(q L)} \, , \label{propMink} 
\ee
where $H^{(1)} = J+i Y$ is the first Hankel function and
$q^2=-k^\mu k_\mu$. For timelike $k$, $q^2> 0$. This expression agrees 
with the scalar Neumann propagators calculated
in~\cite{Garriga:2000yh,Giddings:2000mu} (for both 
points on the brane). Actually, in Minkowski space one can add
a normalizable contribution that modifies this
result~\cite{Balasubramanian:1999sn}. The
Minkowskian propagator that results from Wick rotating the Euclidean
one is the 
one satisfying Hartle-Hawking boundary conditions near the
horizon. The relation of this propagator with the Kaluza-Klein
description of~\cite{Randall:1999vf} is given by the spectral
representation 
\be
\la \varphi(\vec{k}) \varphi(-\vec{k}) \ra = \int_0^\infty \dif \mu
\frac{\sigma(\mu)}{k^2-\mu+i\varepsilon} \,
\ee
Here, $\sigma(\mu)$ is the amplitude at the Planck brane of the wave
function of a Kaluza-Klein mode with mass squared $\mu$. This equation
can be inverted to give
\be
\sigma(\mu)= - \pi {\rm Im} \left[\la \varphi(\vec{k}) \varphi(\vec{-k})
\ra \right]_{q^2=\mu+i\varepsilon} 
\ee
The $\sigma(\mu)$ thus calculated is positive definite. We have
checked numerically that the propagator has no 
poles for timelike or null $\vec{k}$ except in the massless
case. A zero mass corresponds to $\nu=d/2$ and one can use a Bessel
recursion relation to 
show that in this case \cite{Giddings:2000mu}
\be
\la \varphi(\vec{k}) \varphi(\vec{-k}) \ra = - L \left( \frac{d-2}{q^2}
- \frac{1}{q} \frac{H^{(1)}_{d/2-2}(q L)}{H^{(1)}_{d/2-1}(q L)} \right) \, ,
\ee
We have isolated the part with the pole, which is the standard
propagator of a massless scalar field. This corresponds to the zero
mode in the Kaluza-Klein decomposition. The rest, which is suppressed
at low energies by powers of $q/L$, contains no poles and corresponds
to a continuum (with no mass gap) of Kaluza-Klein modes. Scalars with
a positive mass, on the other hand, contain no isolated mode in their
$d$-dimensional description. Nevertheless, we shall see that the
leading term of their propagator in a low energy expansion is a
standard massive $d$-dimensional propagator. Finally, there is an
isolated pole for spacelike $\vec{k}$. Hence, the spectrum contains a
tachyon. This instability has been found in~\cite{Ghoroku:2001pi}
using both the $d+1$ propagator and a Kaluza-Klein approach. The
authors of this reference then argue that the instability might be
cured by the CFT in a holographic interpretation. This does not make
sense since the CFT must give the same effects as the AdS theory. They
also comment that the instability may disappear when the back-reaction
of the scalar on the metric is taken into 
account. The results in~\cite{DeWolfe:2000cp,DeWolfe:2000xi} show that
this is indeed the case.  

In order to compare with the holographic theory we expand the
correlation functions about $\eps=0$. In general we have to
distinguish between fields with integer and with noninteger index
$\nu$, as the expansion of $K_\nu$ contains logarithms for integer
$\nu$. Let us start with the noninteger case. The 1PI
two-point function (\ref{exacttwopoint}) has the expansion
\bea
& \la \varphi(\vec{k}) \varphi(\vec{k^\prime}) \ra^{\rm 1PI}= L^{-1} 
\left(\frac{L}{\eps}\right)^d & \left[ (\nu-\frac{d}{2}) \, + \,
\frac{k^2 \eps^2}{2(\nu-1)} \, \frac{1+\sum_{n=1}^{[\nu]-1}
\frac{(-1)^n \Gamma(\nu-1-n)}{4^n n!\Gamma(\nu-1)} (k\eps)^{2n}}
{1+\sum_{n=1}^{[\nu]-1} \frac{(-1)^n \Gamma(\nu-n)}{4^n n!
\Gamma(\nu)} (k\eps)^{2n}}  \right.  \nn
&& \left.  \mbox{} + \frac{2\Gamma(1-\nu)}{4^\nu\Gamma(\nu)} (k
\eps)^{2\nu} \right] \, +  O(\eps^{2[\nu]-d+2}) \, , \label{ex2pt}
\eea
where $[\nu]$ denotes the entire part of $\nu$ and the sums are
understood to vanish whenever the upper index is smaller than the
lower one. Note that the expansion in $\eps$ is equivalent to a
derivative expansion.
The leading nonlocal part is proportional to
$k^{2\nu}$. After Wick rotation, it gives the leading contribution to
the imaginary part of the Minkowskian two-point function, with the
correct sign. This nonlocal part is the one calculated in standard
AdS/CFT calculations. Since for $\eps=0$ (and $\nu \not \! =
d/2$) the fields $\phi$ diverge or vanish at the real boundary of
AdS, finite correlation functions are obtained by rescaling the
operators in such a way that they do not couple to $\phi(0)$, but
rather to $\lim_{z_0 \rightarrow 0} z_0^{\nu-d/2} \phi(z_0)$, which is
finite. In the regularized AdS/CFT correspondence it is convenient to
normalize the operators in the same way:
\be
\la \O_{\Delta_1} \cdots \O_{\Delta_n} \ra = \eps^{n d -
(\Delta_1+\cdots + \Delta_n)}
\la \varphi_{\Delta_1} \cdots \varphi_{\Delta_n} \ra^{\rm 1PI} \,
. \label{normalization} 
\ee
(For simplicity we ignored this subtlety in
Subsection~\ref{genform}.) Then the leading nonlocal term of (\ref{ex2pt})
gives an $\eps$-independent contribution to correlation functions of
operators. In the standard AdS/CFT ($\eps\rightarrow 0$) this term
yields a conformal finite nonlocal expression, while the local terms
are divergent and must be cancelled by appropriate
counterterms~\cite{Henningson:1998gx,Balasubramanian:1999re,deHaro:2001xn}. 

The expansion~(\ref{ex2pt}) also shows that the leading contribution
to the $\varphi$ propagator is just the usual propagator of a
$d$-dimensional scalar field with squared mass
$(m^\eff)^2=\frac{2(\nu-d/2)(\nu-1)}{\eps^2}$. For $\eps=L$ this
is of the same order as the mass of the bulk field. Both masses are
expected to be of the Planck mass order. The higher-derivative
corrections smooth the behaviour at $k^2\sim -(m^\eff)^2$ and remove the
pole (except in the case $\nu<d/2$, in which the pole remains). So,
for $\nu>d/2$, $\varphi$ does not describe a four-dimensional
particle, but rather a continuous spectral density roughly peaked at
$(m^\eff)^2$. 
It is important to note that boundary mass terms can change
this and give rise to a pole in the propagator. In the Kaluza-Klein
description, the 
reason is that the additional delta functions at the brane can, in
some cases, support a bounded state in the equivalent
quantum-mechanical problem. For $\nu<d/2$ there is an isolated
space-like pole plus a time-like continuum.

From (\ref{ex2pt}) we see that the quadratic part of the effective
action has the form 
\bea
S^\eff & = & \frac{1}{2L} \left(\frac{L}{\eps}\right)^d \int
\dif^d x \, \dif^d y \, \varphi(\vec{x}) 
\left[a_0 \delta(\vec{x}-\vec{y}) + a_1 \eps^2 \Box
\delta(\vec{x}-\vec{y}) + \cdots \right. \nn
&& \left. \mbox{} + a_{[\nu]} \eps^{2[\nu]} \Box^{[\nu]}
\delta(\vec{x}-\vec{y}) 
+ b \eps^{2\nu} \Box^{[\nu]+1}
\frac{1}{|\vec{x}-\vec{y}|^{2(\nu-[\nu]+d/2-1)}} 
+ \cdots \right] \varphi(\vec{y}) \nn
&=& \frac{1}{2L} \int \dif^d x \sqrt{\gind} \varphi(\vec{x}) \,
\left[a_0 \delta(\vec{x}-\vec{y}) + a_1 L^2 \Box_{\rm ind}
\delta(\vec{x}-\vec{y}) + \cdots \right. \nn
&& \left. \mbox{} + a_{[\nu]} L^{2[\nu]} \Box_{\rm ind}^{[\nu]}
\delta(\vec{x}-\vec{y}) + b
L^{2\nu} \Box_{\rm ind}^{[\nu]+1}
\frac{1}{\left((x^i-y^i)(x_i-y_i)\right)^{(\nu-[\nu]+d/2-1)}}   
+ \cdots \right] \varphi(\vec{x}) \, , \nn
\mbox{}
\eea
where $x^i x_i = g^{\mbox{\tiny ind}}_{ij} x^i x^j$ and $\Box_{\rm
ind}= g_{\mbox{\tiny ind}}^{ij}\d_i\d_j$. The nonlocal term is 
nothing but a renormalized expression of
$|\vec{x}-\vec{y}|^{-2\Delta}$. The 
regularization and renormalization of these expressions is discussed
in Section~\ref{CFT} 
and some useful formulae are collected in the Appendix. 
As emphasized in~\cite{deHaro:2001xn}, the effective action does not
depend explicitly on $\eps$ when expressed in terms of the induced
metric and scalar field. This gives the physical meaning of the
expansion in $\eps$ for an observer living on the brane: it corresponds to a
low-energy expansion with energies measured in units of
$L^{-1}$. Note that we can rescale the brane coordinates such that
$\gind=\delta$ (equivalently, $\eps=L$).

Consider now an integer $\nu$. For $\nu \geq 1$ the expansion of the
two-point function is the same as above, but with the
$\eps^{2(\nu-d/2)}$ term given by
\be
\frac{(-1)^\nu 4^{1-\nu}}{\Gamma(\nu)^2} k^{2\nu}\eps^{2(\nu-d/2)}
\ln(k\eps) + c \, k^{2\nu} \eps^{2(\nu-d/2)} \, ,
\ee
with $c$ a constant number.
In the renormalized AdS/CFT correspondence, this logarithm has to be
cancelled by a counterterm proportional to $\log{M \eps}$, with $M$ an
arbitrary scale. This produces an anomalous breaking of conformal
invariance in the CFT which agrees with the violation of
conformal invariance usually introduced in the renormalization
procedure~\cite{Henningson:1998gx,Erdmenger:2001ja}. The effective
mass has the 
same expression as in the noninteger case. It vanishes
if and only if the bulk field is massless ($\nu=d/2$). In this case,
the pole at $k^2=0$ is obviously preserved by the corrections and
$\varphi$ contains a discrete mode.

The case $\nu=0$ corresponds to the smallest mass allowed by
unitarity, $m^2=-d^2/4$,
and has some special features. The two-point function reads in this
case
\be
\la \varphi(\vec{k}) \varphi(\vec{-k}) \ra^{\rm 1PI}=
L^{-1} \left( \frac{L}{\eps} \right)^d \,
\left[-\frac{d}{2}-\frac{1}{\ln (k \eps \gamma_E / 2)}\right] \,
\label{twopointnu0} 
\ee
where $\gamma_E=1.781\ldots$ is the Euler constant. The appearance of
a logarithm in the denominator is related to the fact that the
bulk field diverges as $z_0^{d/2} \ln(z_0 k)$ near the AdS
boundary. 
It seems reasonable to define the correlation function of
two operators of conformal dimension $d/2$ by \cite{Minces:2000eg}
\be
\la \O(\vec{k}) \O(-\vec{k}) \ra= \left(\eps^{\frac{d}{2}} \ln (\eps
M^\prime)\right)^2 \, \la \varphi(\vec{k}) \varphi(\vec{-k}) \ra^{\rm
1PI} \, \label{raro}
\ee
for some scale $M^\prime$. $\la \O \O \ra$ then contains the correct
finite nonlocal term of the CFT, $\ln(k M^\prime)$, with corrections
of order $1/\ln(\eps M^\prime)$ that vanish logarithmically as
$\eps\rightarrow 0$. In the local terms, on the other hand,
(divergent) double
logarithms appear. This is not expected from the CFT analysis in
Section~\ref{CFT}. Furthermore, we have found that the extension of
(\ref{raro}) to three-point functions does not even give the correct
nonlocal part. Therefore, the regularized AdS/CFT needs some
modification in this particular case. This may be related to the
fact that the Dirichlet propagator defined in (\ref{Dprop}) cannot
describe fields with dimension $(d-2)/2\leq \Delta <d/2$. Indeed, for
a given 
mass, this propagator automatically selects the largest conformal
dimension ($\Delta$ and not $\Delta_-$). $\nu=0$ is the special case
where $\Delta=\Delta_-$. We will not attempt to find a correct
prescription for $\Delta=d/2$ or $\Delta<d/2$ in this paper, but
leave it as an interesting open problem\footnote{This is only a
problem of the cutoff regularization (which is the relevant one for
Randall-Sundrum). The renormalization procedure proposed
in~\cite{Klebanov:1999tb} gives correct CFT correlations functions for
$\Delta\leq d/2$.}.

%%%%%%%%%%%%%%%

\subsection{Three-point functions}

Suppose now that the action contains a cubic term of the form
\be
S \supset \frac{\lambda_{ijk}}{3!} \, \int_\eps \dif^{d+1} z \, \sqrt{g} \,
\phi_i \phi_j \phi_k \, , \label{cubicaction}
\ee
where $\phi_i$ is a scalar field of dimension $\Delta_i$ and the
coupling $\lambda_{ijk}$ is completely symmetric in the indices. In
momentum space, the 1PI three-point functions in the effective theory
read 
\be
\la \varphi_1(\vec{k}_1) \varphi_2(\vec{k}_2) \varphi_3(\vec{k}_3)
\ra^{\rm 1PI} = - \lambda_{123} \int_\eps^{\infty} \dif z_0 \,
\left(\frac{L}{z_0}\right)^{d+1} \prop_{\nu_1}(z_0,\vec{k}_1)
\prop_{\nu_2}(z_0,\vec{k}_2) \prop_{\nu_3}(z_0,\vec{k}_3) \, .  
\ee
The integral over $z_0$ is difficult to perform in
general. Fortunately, it is possible to compute in a simple manner
the leading terms in the $\eps$ expansion. We distinguish local terms,
for which all points are coincident (two delta functions in coordinate
space), semilocal terms, for which two points coincide and the other
is kept appart (one delta function), and completely nonlocal terms,
for which all points are noncoincident. We shall calculate all the
terms up to the leading completely nonlocal term. First, observe
that the integral has the form 
\be
\la \varphi_1(\vec{k}_1) \varphi_2(\vec{k}_2) \varphi_3(\vec{k}_3)
\ra^{\rm 1PI} = - \lambda_{123} L^{d+1} \frac{G(\eps)}{f(\eps)} \, ,
\label{3ptform} 
\ee
with
\be
G(\eps)=\int_\eps^\infty \dif z_0\, z_0^{-d-1} f(z_0) \,
. \label{Gdef} 
\ee
Consider noninteger indices $\nu_1$, $\nu_2$ and $\nu_3$.
The function $f(z_0)$ can be expanded about $z_0=0$:
\bea
f(z_0) & = & \prod_{i=1}^3 \prop_{\nu_i}^0(z_0,\vec{k}_i) \nn
& = & C_\nu \prod_{i=1}^3 z_0^{d/2-\nu_i} \left(\sum_{n=0}^\infty a_n^{(i)}
z_0^{2n} + \beta^{(i)} z_0^{2\nu_i} \sum_{n=0}^\infty b_n^{(i)}
z_0^{2n} \right) \, ,  \label{fexpansion}
\eea
where
\be
\prop_{\nu}^0(z_0,\vec{k})  =  C_\nu k^\nu z_0^{d/2} K_\nu(z_0 k)
\ee
is the stantard $\eps=0$ bulk-to-boundary propagator,
$C_\nu=\frac{\sin \pi\nu}{2^{\nu-1}\pi} \Gamma(1-\nu)$, and the explicit
expressions of the momentum-dependent coefficients are
\bea 
a_n^{(i)} &=& \frac{\Gamma(1-\nu_i)}{n!\Gamma(1-\nu_i+n)}
\left(\frac{k_i}{2}\right)^{2n} \, , \nn
b_n^{(i)} &=& \frac{\Gamma(1+\nu_i)}{n!\Gamma(1+\nu_i+n)}
\left(\frac{k_i}{2}\right)^{2n} \, , \nn
\beta^{(i)} &=& - \frac{\Gamma(1-\nu_i)}{n!\Gamma(1+\nu_i)}
\left(\frac{k_i}{2}\right)^{2\nu_i} \, .
\eea
Note that $a_0^{(i)}=b_0^{(i)}=1$. Using (\ref{fexpansion}) in
(\ref{Gdef}),
\bea
\lefteqn{G(\eps)= C_\nu \int_\eps^\infty \dif z_0 z_0^{d/2-1-\sigma} \left[
\sum_{n=0}^{[\frac{1}{2}(\sigma-\frac{d}{2})]} \tilde{a}_n z_0^{2n} +
\sum_{i=1}^3 \left(\beta^{(i)} z_0^{2\nu_i}
\sum_{n=0}^{[\frac{1}{2}(\sigma-\frac{d}{2})-\nu_i]} \tilde{a}_n^{(i)}
z_0^{2n}\right) \right. } && \nn
&& \mbox{} \left. + \sum_{i\not \! = j=1}^3
\left(\beta^{(i)}\beta^{(j)} z_0^{2(\nu_i+\nu_j)}
\sum_{n=0}^{[\frac{1}{2}(\sigma-\frac{d}{2})-\nu_i-\nu_j]}
\tilde{a}_n^{(i,j)} z_0^{2n} \right) \right] \, +  \tilde{G}(\eps) \,
,  \label{Gintexp}
\eea
where $\sigma=\sum_{i=1}^3 \nu_i$ and
\bea
\tilde{a}_n & = & \sum_{n_1=0}^n \sum_{n_2=0}^{n-n_1}
\sum_{n_3=0}^{n-n_1-n_2} a^{(1)}_{n_1} a^{(2)}_{n_2} a^{(3)}_{n_3} \,
, \nn
\tilde{a}_n^{(1)} & = & \sum_{n_1=0}^n \sum_{n_2=0}^{n-n_1}
\sum_{n_3=0}^{n-n_1-n_2} b^{(1)}_{n_1} a^{(2)}_{n_2} a^{(3)}_{n_3} \,
, \nn
\tilde{a}_n^{(1,2)} & = & \sum_{n_1=0}^n \sum_{n_2=0}^{n-n_1}
\sum_{n_3=0}^{n-n_1-n_2} b^{(1)}_{n_1} b^{(2)}_{n_2} a^{(3)}_{n_3} \,
.
\eea
Other $\tilde{a}_n^{(i)}$ and $\tilde{a}_n^{(i,j)}$ are analogously
defined. The rest $\tilde{G}$ is at most logarithmically 
divergent when $\eps\rightarrow 0$. Let us assume momentarily that the
conformal dimensions of the fields are such that
$\frac{1}{2}(\sigma-\frac{d}{2})$,
$\frac{1}{2}(\sigma-\frac{d}{2}-\nu_i)$, and 
$\frac{1}{2}(\sigma-\frac{d}{2}-\nu_i-\nu_j)$ are neither positive
integers nor zero. Then, each of the terms that we have written explicitly
in (\ref{Gintexp}) vanishes sufficiently rapidly at infinity, and we can
perform the corresponding integrals. Furthermore, in this case
$\tilde{G}(\eps)$ is finite as $\eps \rightarrow 0$. We find
\be
G(\eps)= G^{\rm div}(\eps) + \tilde{G}(0) + O(\eps^\eta) \, , \label{mini}
\ee
with $\eta>0$ and
\bea
\lefteqn{G^{\rm div}(\eps) = C_\nu
\frac{\eps^{d/2-\sigma}}{\frac{d}{2}-\sigma} \left[ 
\sum_{n=0}^{[\frac{1}{2}(\sigma-\frac{d}{2})]}
\frac{(\frac{d}{2}-\sigma) \tilde{a}_n}{\frac{d}{2}-\sigma+2n}
\eps^{2n} \right. } && \nn 
&&  \mbox{}  +
\sum_{i=1}^3 \left(\beta^{(i)} \eps^{2\nu_i}
\sum_{n=0}^{[\frac{1}{2}(\sigma-\frac{d}{2})-\nu_i]}
\frac{(\frac{d}{2}-\sigma) \tilde{a}_n^{(i)}}{\frac{d}{2}-\sigma+\nu_i+2n}
\eps^{2n}\right)  \nn
&& \mbox{} \left. + \sum_{i \not \! = j=1}^3 \left(\beta^{(i)}\beta^{(j)}
\eps^{2(\nu_i+\nu_j)}
\sum_{n=0}^{[\frac{1}{2}(\sigma-\frac{d}{2})-\nu_i-\nu_j]}
\frac{(\frac{d}{2}-\sigma)
\tilde{a}_n^{(i,j)}}{\frac{d}{2}-\sigma+2\nu_i+2\nu_j+2n} z_0^{2n}
\right) \right] \, ,  \label{Gexp}
\eea
The 1PI three-point function is obtained by using (\ref{mini}),
(\ref{Gexp}) and (\ref{fexpansion}) in (\ref{3ptform}). It has the
form
\bea
&& \eps^{-d} \left[ \sum_n \alpha_n \eps^{2n}
+ \sum_i  \beta^{(i)} \sum_n \alpha_n^{(i)} \eps^{2(\nu_i+n)} +
\sum_{i,j}  \beta^{(i)} \beta^{(j)} \sum_n \alpha_n^{(i,j)}
\eps^{2(\nu_i+\nu_j+n)} \right]   \nn
&& \mbox{} + \eps^{\sigma-3\frac{d}{2}} \tilde{G}(0) +
O(\eps^{\sigma-3\frac{d}{2}+\eta}) \, .  
\eea
All the terms inside the bracket are either local or semilocal, except
the ones with $\beta^{(i)} \beta^{(j)}$, $i\not \! = j$, which are
proportional to 
$k_i^{2\nu_i} k_j^{2\nu_j}$. To order $\eps^{\sigma-3\frac{d}{2}}$
these terms only contribute when $\sigma>d/2-\nu_i-\nu_j$ (remember
that we have assumed $\sigma \not \! = d/2-\nu_i-\nu_j$ for the
moment), \ie, when one of the conformal 
dimensions, $\Delta_i$, is greater than the sum of the others. The
coefficients $\lambda_{ijk}$ of such ``superextremal'' functions
vanish in type IIB supergravity on $AdS_5$ compactified on $S^5$, due
to the properties of the $S^5$ spherical harmonics. (In the dual ${\cal
N}=4$ theory these functions are forbidden by R symmetry.) More
generally, for any theory on $AdS_{d+1}$, the completely nonlocal part
of superextremal $n$-point functions diverges---after the rescaling 
(\ref{normalization})---when $\eps\rightarrow 0$. This
invalidates the (unregulated) AdS/CFT correspondence unless
$\lambda_{ijk}=0$ when $\Delta_i > \Delta_j+\Delta_k$. For these
reasons, we assume that $\lambda_{ijk}$ does vanish in any
superextremal case. On the other hand, the terms with one
$\beta^{(i)}$ are semilocal. These terms are
relevant in our approximation when $\sigma>d/2+2\nu_i$, and in this
case are divergent after rescaling when $\eps\rightarrow 0$. One might
be worried about the fact that one needs to add nonlocal counterterms
to $S^\eff$ in order to obtain a finite renormalized effective
action. However, this is not so strange since $S^\eff$ itself is
nonlocal. Presumably, the nonlocal counterterms arise from local
counterterms in the original action when the bulk degrees of freedom
are integrated out. The holographic counterpart of this
renormalization procedure is discussed in Section~\ref{CFT}. 

The leading completely nonlocal term is given by $\tilde{G}(0)$. It
can be calculated in coordinate space using 
conformal techniques~\cite{Freedman:1999tz}. The result is
\be
\frac{c}{x_{12}^{\Delta_1+\Delta_2-\Delta_3}
x_{13}^{\Delta_1+\Delta_3-\Delta_2}
x_{23}^{\Delta_2+\Delta_3-\Delta_1}} \, , \label{conf3pt}
\ee
with
\bea
&& c=- \frac{\Gamma[\frac{1}{2}(\Delta_1+\Delta_2-\Delta_3)] 
\Gamma[\frac{1}{2}(\Delta_1+\Delta_3-\Delta_2)]
\Gamma[\frac{1}{2}(\Delta_2+\Delta_3-\Delta_2)] }{2\pi^d
\Gamma[\Delta_1-\frac{d}{2}] \Gamma[\Delta_1-\frac{d}{2}]
\Gamma[\Delta_1-\frac{d}{2}]} \nn
&& \mbox{} \times \Gamma[\frac{1}{2}(\Delta_1+\Delta_2+\Delta_3-d)] \,
. 
\eea
We have defined $x_{ij}=|\vec{x}_i - \vec{x}_j|$.
The functional form in (\ref{conf3pt}) is dictated by conformal
invariance. As it stands, this expression is valid only for
noncoincident points. $\tilde{G}(0)$ is given by its renormalized value. 
In order to find the three-point functions of CFT operators we must
rescale according to (\ref{normalization}).

We have found the 1PI three-point function to order
$\eps^{\sigma-3\frac{d}{2}}$, at least in coordinate 
space and excluding some particular values of the conformal
dimensions. These exceptions are interesting. They correspond to
either the presence of logarithms in the expansion of $f(z_0)$ (when
at least one $\nu_i$ is integer) or to $1/z_0$ terms in the integrand
of (\ref{Gintexp}), which also give rise to logarithms. In the first
case, it is straightforward to modify (\ref{Gexp}) to incorporate the
logarithms in the expansion of $f(z_0)$. Let us see how to deal with
the second possibility. The integral of the $1/z_0$ terms
diverges at infinity and cannot be performed independently. 
Instead, we can split the logarithmically divergent integrals in the
following way:
\be
\int_\eps^\infty \dif z_0\, z_0^{-1} = - \ln (M\eps) +
\int_{M^{-1}}^\infty \dif z_0\, z_0^{-1} \, . \label{splitlog}
\ee
The second term can then be included in $\tilde{G}(\eps)$, which is a
well-defined integral because the sum of all terms is well-behaved at
infinity. Moreover, $\tilde{G}(\eps)$ is convergent for
$\eps\rightarrow 0$ and can be written as
$\tilde{G}(0)+O(\eps^\eta)$. Since the complete $G(\eps)$ does not
depend on the arbitrary scale $M$, $\tilde{G}(0)$ must contain terms
of the form $\ln (k/M)$. Again, $\tilde{G}(0)$ can be obtained from
an standard $\eps=0$ AdS/CFT calculation. As discuss in the next
section, the logarithms arise naturally when the conformal result is
renormalized. Higher powers of logarithms appear when
one or more indices are integer and the integrand contains
terms of the form $\ln(k z_0)/z_0$. 

In all these cases it is possible that the logarithms do not appear to
order $\eps^{\sigma-3\frac{d}{2}}$.  
Let us enumerate the special possibilities ($n$ is a positive
or vanishing integer): 
\begin{enumerate}
\item $\sigma=d/2+2n$. Logarithms appear at order
$\eps^{\sigma-3d/2}$.
\item $\sigma=d/2+2\nu_i+2n$. Logarithms multiplying $k_i^{2\nu_i}$
appear at order $\eps^{\sigma-3d/2}$. 
\item $\sigma=d/2+2\nu_i+2\nu_j$. This case has very special features
and we discuss it below.  
\item $\nu_i$ integer and $\sigma\geq d/2+2\nu_i$. Logarithms
multiplying $k_i^{2\nu_i}$ appear at order 
$\eps^{2\nu_i + 2n-d}$, $n=0,1,\ldots,[\frac{1}{2}(\sigma-3d/2)]$.  
\end{enumerate}
Combinations of these possibilities give rise to higher powers of
logarithms. 

When $\sigma=d/2+2\nu_i+2\nu_j$, the conformal
dimension of one of the fields equals the sum of the remaining
dimensions. In this ``extremal'' case, a $\ln \eps$ appears in the
leading completely nonlocal term. Therefore, the rescaled correlation
function for the 
CFT diverges logarithmically as $\eps\rightarrow 0$. The divergence
arises from the region near the boundary point where the field with
highest dimension is inserted. It seems again
that the coupling should vanish in a consistent gravity theory
with an ($\eps=0$) CFT dual. However, it is
known that the extremal correlators of chiral primary
operators in ${\cal N}=4$ SYM have a nonvanishing finite
expression. The solution to this problem was given
in~\cite{D'Hoker:1999ea}: the supergravity theory contains couplings
with 
different number of derivatives, such that the divergence and all the
bulk contributions cancel and only surface terms
remain.\footnote{Usually, one performs field redefinitions
to get rid of the higher derivative couplings. It turns out that
the resulting extremal couplings vanish, but one must include
surface terms when performing the field
redefinitions~\cite{Arutyunov:2000im}. Alternatively, one can continue
analitically the conformal
dimensions; then the pole in the integral in the region near the
boundary is cancelled by a zero in the coupling. Again, a finite
result is obtained~\cite{D'Hoker:1999ea,D'Hoker:2000dm}.} In
fact, the very same mechanism takes place for two-point functions,
which are the simplest example of extremal correlators. In that case,
the contributions of the mass and the kinetic terms are also
separately divergent, but they cancel out leaving only the surface
contribution (\ref{exacttwopoint})\footnote{As in any
$\infty$-$\infty$ situation one has to regularize to find a definite
answer. D.Z. Freedman \etal\ used the cutoff regularization and found a
result that agrees with the Ward identity relating the two-point
function of two scalars and the three-point function of two scalars and
a current~\cite{Freedman:1999tz}. Since the latter is power-counting
finite for noncoincident points, it follows that the normalization
obtained in~\cite{Freedman:1999tz} is universal for any regularization
preserving the gauge 
symmetry for noncoincident points. Analytical continuation also
respects the gauge symmetries but it cannot be used for the particular
case of the two-point function (since one cannot avoid going into 
a superextremal situation).}.
The correlation functions of extremal $n$-point functions
have been well studied both in supergravity theories and in the dual
superconformal
theories~\cite{D'Hoker:1999ea,Bianchi:1999ie,Eden:2000kw,Eden:2000gg}.
It has been shown that the field  
theory extremal correlators obey nonrenormalization theorems, in the
sense that their value is independent of the coupling. This is related
to the fact that extremal correlators factorize into a product of
two-point functions. It has also been shown that next-to-extremal
functions are neither
renormalized~\cite{Eden:2000kw,Erdmenger:2000pz,Eden:2000gg} 
and, more generally, that ``near-extremal'' functions satisfy special
factorization properties~\cite{D'Hoker:2000dm}. This strongly suggests a
generalized 
consistent truncation for type IIB supergravity on $AdS_5\times
S^5$~\cite{D'Hoker:2000dm}, that agrees with explicit calculations of
cubic~\cite{Lee:1998bx,Arutyunov:2000en,Corrado:1999pi} and
quartic~\cite{Arutyunov:2000fb} couplings. These
arguments have also been used to conjecture  
generalized consistent truncation in type 
IIB supergravity on $AdS_{(4|7)}\times S^{(7|4)}$, for which the CFTs are
not so well known~\cite{D'Hoker:2000vb}. This agrees with the cubic couplings
in~\cite{Corrado:1999pi,Bastianelli:1999en,Bastianelli:2000vm}.
It is plausible that the gravity theory relevant to Randall-Sundrum
also has these properties. In the following we assume that this is
the case at least for extremal functions (which would otherwise either
vanish or diverge for $\eps \rightarrow 0$).

Let us come back to the finite $\eps$ AdS/CFT.  From the assumption
above, our Randall-Sundrum gravity theory contains cubic couplings
with different number of derivatives that 
conspire to render the completely nonlocal part of extremal
three-point functions finite. The extremal three-point functions then
reduce to surface terms and can be calculated exactly. As an example, we
consider type IIB supergravity and the correlator studied
in~\cite{D'Hoker:1999ea}, $\la \varphi_{\Delta_1}\varphi_{\Delta_2} 
s_{\Delta_1+\Delta_2}\ra^{\rm 1PI}$. $\varphi$ and $s$ are 
boundary values of Kaluza-Klein modes (from the $S^5$ reduction) of
the dilaton and of a mixture of the 
4-form and the graviton (with indices on the sphere). Integrating
by parts and using of the equations of motion, this correlator is
reduced to the surface term
\be
\eps^{-d+3} \lim_{z_0\rightarrow \eps} \left[\d_0
\prop_{\nu_1}(z_0,\vec{k_1}) \d_0 \prop_{\nu_2}(z_0,\vec{k_2}) \d_0
\prop_{\nu_1+\nu_2+d/2}(z_0,\vec{k_3}) \right] \, . \label{extremal3pt}
\ee
The leading nonlocal part was found in~\cite{D'Hoker:1999ea} to have
the factorized structure
\be
\eps^{2(\nu_1+\nu_2)-3d/2} \, [k_1^{2\nu_1} \ln(k_1 \eps)] \, [
k_2^{2\nu_2} \ln(k_2\eps)] \, . 
\ee
The logarithms only appear for integer $\nu_i$. This is the Fourier
transform of the product of two (renormalized) two-point
functions. Without further effort we can compute the local and
semilocal terms in (\ref{extremal3pt}) as well, and see whether this
factorization property holds for the complete regularized functions. We
find that it does: the regularized extremal three-point function is,
to order $\eps^{2(\nu_1+\nu_2-3d/2)}$, a product of two-point functions:
\be
\la \varphi_{\nu_1} (\vec{k}_1)\varphi_{\nu_2}(\vec{k}_2)
s_{\nu_1+\nu_2+d/2}(\vec{k}_3) \ra^{\rm 1PI} \propto \la
\varphi_{\nu_1}(\vec{k}_1) \varphi_{\nu_1}(-\vec{k}_1) \ra^{\rm 1PI}
\la \varphi_{\nu_2}(\vec{k}_2)
\varphi_{\nu_2}(-\vec{k}_2) \ra  ^{\rm 1PI}
\ee
It is reasonable to expect that the factorization of
regularized extremal correlation functions into products of two-point
functions holds to this order in any gravity theory with
generalized consistent truncation. On the other hand, as we mentioned
above, we have found that when some of the fields have dimension $d/2$, the
generalization of the prescription (\ref{raro}) does not lead to a
correct nonlocal part of the three-point correlator (at least in the
extremal case). This deserves further study.

%%%%%%%%%%%%%%%%%%%%%%%%%%%%%%%%%%%%%%%%%%%%%%%%%%%%%%

\section{The holographic description of RS2}
\label{CFT}

\subsection{Regularization and renormalization of the conformal
theory}

Any CFT has vanishing beta functions, \ie, contains no
divergencies. However, in the AdS/CFT correspondence one does not
consider the pure CFT but its deformations by composite operators built
out of the elementary field of the theory. The insertions of operators
introduce UV divergencies that must be cancelled by local
counterterms. Consider two-point correlation functions in
a $d$-dimensional CFT,
\bea
\la \O_\Delta(x) \O_\Delta(0) \ra & = & \left[\frac{\delta
W^\CFT[\varphi]}{\delta \varphi_\Delta(x)
\delta \varphi_\Delta(0)}\right]_{\varphi=0} \nn
& \propto & \frac{1}{x^{2\Delta}} \, .
\eea
Since in this section all the vectors are $d$-dimensional, we use $x$
to indicate both a vector and its Euclidean modulus.
$\Delta$ is the conformal dimension of the operator and
$\varphi_\Delta$ is the source coupled to the operator in the path
integral (\ref{WCFT}). For $\Delta \geq d/2$ this expression is too
singular at coincident points ($x=0$) to behave as a tempered
distribution. In other words, its Fourier transform is
divergent. To make sense of it one must renormalize the
deformed CFT. The counterterms one needs for this function
have the form 
\be
S^{\CFT\, {\rm ct}} \supset \int \dif^d x \varphi_\Delta(x) Q_{\Delta,\Delta}
\varphi_\Delta(x) \, \label{quadcounterterms} 
\ee
where $Q_{\Delta,\Delta} = c_0 + c_1 \Box + \ldots$ is a differential
operator of mass dimension $2\Delta$. We can for instance use
dimensional regularization and minimal substraction to find the
renormalized expression~\cite{Dunne:1992ws,delAguila:1999bd}
\be
\la \O_\Delta(x) \O_\Delta(0) \ra^R \propto \Box^{[\Delta-d/2]+1}
\frac{1}{x^{2(\Delta-[\Delta-d/2]-1)}} \label{ex1}
\ee
for noninteger $\Delta-d/2$ and
\be
\la \O_\Delta(x) \O_\Delta(0) \ra^R \propto \Box^{\Delta-d/2+1}
\frac{\ln (x M)}{x^{d-2}} \label{ex2}
\ee
for integer $\Delta-d/2$. In latter case one cannot avoid
introducing a dimensionful scale $M$.
Since dimensional regularization only detects logarithmic
divergencies, the counterterm operator is simply
$Q_{2\Delta}=0$ for noninteger $\Delta-d/2$, and  $Q_{2\Delta}\propto
\frac{1}{\varepsilon} \Box^{\Delta-d/2}$ for integer $\Delta-d/2$.
In both renormalized expressions, the derivatives are prescribed to
act by parts on test functions, as in differential
renormalization~\cite{Freedman:1992tk}. 
A change of renormalization scheme would only modify the renormalized
result by local terms.

Three-point functions have both subdivergencies and overall
divergencies. The former can be cancelled by a local counterterm
with two $\varphi$ coupled to one CFT operator, and the latter by a
local counterterm cubic in $\varphi$. The local counterterms in the
action generate nonlocal counterterms for the generating functional
when the CFT are integrated out. Since the pure CFT is
finite, we only need the counterterms that correct the singular
behaviour when external points coincide. That is to say, we need not
worry about internal points in a Feynman diagram description of the
correlators. If we wanted to calculate correlation functions involving
both operators and elementary fields of the CFT, $A$, we would also
need counterterms mixing $\varphi$ and $A$. As far as we know, the 
renormalization of the CFT in the context of  AdS/CFT has only
been studied in~\cite{Chalmers:2000gc}. We believe that this is an important
aspect of the correspondence that deserves further study. Here we are
mainly interested in the CFT at the regularized level. Different
regularizations will be dual to 
different regularizations of the AdS theory and we are interested in
the regularization relevant for Randall-Sundrum. The only thing we know
about this regularization is that the regulator is a dimensionful
parameter (which excludes dimensional regularization). Nevertheless,
many of the features we found in AdS can be shown on general grounds
to have its counterpart in the regularized CFT.

It will be convenient to use
differential regularization~\cite{Freedman:1992tk}\footnote{This
method can be understood either as a regularization or as a renormalization
procedure, and we shall use the terms ``differential regularization''
and ``differential renormalization'' to distinguish both
interpretations.} This method is well suited for conformal theories as
it naturally works in coordinate space and only modifies correlation
functions at coincident
points~\cite{Freedman:1992tz,Osborn:1994cr,Erdmenger:1997yc}.  
The idea of differential regularization and renormalization is simple:
substitute expressions that are too singular at 
coincident points by derivatives of well-behaved distributions, such
that the original and the modified expression are identical at
noncoincident points. The derivatives are prescribed to act by parts
on test functions, disregarding divergent surface terms around the
singularity. For instance, for $d=4$, 
\be
\frac{1}{x^6} \rightarrow \frac{-1}{32} \Box\Box \frac{\ln (xM)}{x^2} +
a_0 \mu^2 \delta(x) + a_1 \Box \delta(x) \, .  \label{DRexample}
\ee
In a Fourier transform the total derivatives just give
powers of momenta. Since both sides of (\ref{DRexample})
must only be equal for nonzero $x$, the local terms are
arbitrary. Note that changing the mass scale $M$, which is required to
make the argument of the logarithm dimensionless, is equivalent to a
redefinition of $a_1$. This is related to renormalization group
invariance. Although the \rhs\ 
of (\ref{DRexample}) is a correct renormalized expression, here we
want to interpret the \rhs\ of (\ref{DRexample}) as 
the regularized value of the \lhs. Since on the AdS side there is one
single dimensionful parameter, we take $M=\mu=1/\eps$ and consider
$\eps k \ll 1$ for any momentum $k$ in the Fourier transformed
functions. Note that in an arbitrary regularization we could add more
local and nonlocal terms 
that vanish as $\eps\rightarrow 0$. They would correspond to higher
order terms that do not appear in the
approximation we used in the AdS calculations. Some useful differential
identities and Fourier transforms are collected in the Appendix. 

\subsection{Two-point functions}

The two-point function of two scalar operators of dimension $\Delta$ is
determined, for noncoincident points, by conformal invariance:
\be
\la \O_\Delta(x) \O_\Delta(0) \ra = N \frac{1}{x^{2\Delta}} \, ,
~x\not \! = 0 \, .
\ee
$N$ is a constant depending on the normalization of the
operators. We shall fix it below such that it agrees with the AdS
result. In order to extend this expression to a well-defined
distribution over all space, we use differential regularization. 
Again, we have to distinguish two cases: integer
$\nu=\Delta-d/2$ and 
noninteger $\nu$. For noninteger $\nu$ the regularized expression is
\bea
\la \O_\Delta(x) \O_\Delta(0) \ra = N  [ C_\Delta \Box^{[\nu]+1}
\frac{1}{x^{2(\Delta-[\nu]-1)}} + a_0 \eps^{-2\nu} \delta(x)  \nn
\mbox{} + a_1 \eps^{-2(\nu-1)} \Box \delta(x) + \cdots + a_{[\nu]}
\eps^{-2(\nu-[\nu])} \Box^{[\nu]} \Delta(x)] \, ,
\eea
with 
\be
C_\Delta= \frac{\Gamma(\Delta-[\nu]-1) \Gamma(\nu-[\nu])}{4^{[\nu]+1}
\Gamma(\Delta)\Gamma(\nu+1)} \, .
\ee
This is a well-defined distribution when the derivatives act ``by
parts''. Using the relation (\ref{normalization}), a convenient 
normalization $N$ and the Fourier transforms in the Appendix, we find 
\be
\la \varphi_\nu(k) \varphi_\nu (-k) \ra^{\rm 1PI} = L^{-1}
\left(\frac{L}{\eps}\right)^d \left[ \bar{a}_0 - \bar{a}_1 k^2 \eps^2
+ \cdots + (-1)^{[\nu]} \bar{a}_{[\nu]} k^{2[\nu]} \eps^{2[\nu]} +
\frac{2\Gamma(1-\nu)}{4^\nu \Gamma(\nu)} k^{2\nu} \eps^{2\nu} \right]
\, ,
\ee
which agrees with the structure in (\ref{ex2pt}) to order
$\eps^{2\nu-d}$. We have defined
\be
\bar{a}_i= - \frac{2\nu \Gamma(\Delta)}{\pi^{d/2} \Gamma(\nu)} a_i
\ee
Exact agreement is found only for a particular regularization, \ie,
for particular values of $a_i$: 
\bea
\bar{a}_0 & = & \Delta \, , \nn
\bar{a}_1 & = & \frac{1}{2(1-\nu)} \, , \nn
\ldots
\eea

For integer $\nu$ we find
\bea
\la \O_\Delta(x) \O_\Delta(0) \ra = N  [ C^\prime_\Delta \Box^{\nu+1}
\frac{\ln (x/\eps)}{x^{d-2}} + a_0 \eps^{-2\nu} \delta(x)  \nn
\mbox{} + a_1 \eps^{-2(\nu-1)} \Box \delta(x) + \cdots + a_{\nu}
\Box^{\nu} \delta(x)] \, ,
\eea
with
\be
C^\prime_\Delta= \frac{\Gamma(\frac{d}{2})} {4^\nu (2-d)
\Gamma(\Delta)\Gamma(\nu+1)} \, .
\ee
The logarithm appears when writing the unregularized expression as
a total derivative (see the Appendix). Again, this agrees with the AdS
calculation for particular values of $a_i$. 

To obtain renormalized two-point functions, one would need counterterms
of the form (\ref{quadcounterterms}). In a minimal substraction
scheme, 
\bea
Q_{2\Delta} & = & - N \left( a_0 \eps^{-2\nu}
+ a_1 \eps^{-2(\nu-1)} \Box + \cdots + a_{[\nu]}
\eps^{-2(\nu-[\nu])} \Box^{[\nu]}]  \right. \nn
&& \left. \mbox{} + C^\prime_\Delta
\frac{4\pi^{d/2}}{\Gamma(\frac{d}{2}-1)} \ln(M\eps) \Box^{\nu+1}
\right) \, ,
\eea
where the last term only appears if $\nu$ is integer, and $M$ is the
renormalization scale. The renormalized two-point
function in this scheme reads
\be
\la \O_\Delta(x) \O_\Delta(0) \ra_R = N C_\Delta \Box^{[\nu]+1}
\frac{1}{x^{2(\Delta-[\nu]-1)}}
\ee
and
\be
\la \O_\Delta(x) \O_\Delta(0) \ra_R = N C^\prime_\Delta \Box^{\nu+1}
\frac{\ln (x M)}{x^{d-2}} \, ,
\ee
for noninteger and integer $\nu$, respectively.

\subsection{Three-point functions}

The structure of regularized three-point functions is much richer. The
reason is that divergencies can appear when the three 
points are coincident (overall divergence) or when two points coincide
and the other is kept apart (subdivergence). For noncoincident points,
the correlation 
functions of three scalar operators are completely determined by
conformal invariance (up to normalization):
\bea
\la \O_{\Delta_1}(x_1) \O_{\Delta_2}(x_2) \O_{\Delta_3}(x_3) \ra & = & N
\frac{1}{x_{13}^{\Delta_1+\Delta_3-\Delta_2}
x_{23}^{\Delta_2+\Delta_3-\Delta_1}
x_{12}^{\Delta_1+\Delta_2-\Delta_3}} \nn
&=& N \frac{1}{x^{\omega_2}
y^{\omega_1}
(x-y)^{\omega_3}} \, .
\eea
In the second identity we have changed variables to $x=x_{13}$,
$y=x_{23}$, $z=x_3$ ($z$ does not appear due to translation
invariance) and defined $\omega_i$ as linear combinations of the
conformal dimensions. In general, this function can be regularized with 
differential regularization using the systematic procedure developed
in~\cite{Latorre:1994xh}. Here we shall just describe some features of
the regularized functions and compare with the corresponding AdS
correlation functions. A particularly simple example is worked out in
detail in the Appendix.

First, we observe that semilocal terms appear in the CFT
description when subdivergencies are regularized. There are
subdivergencies when, for some $i$, $\omega_i\geq d$. Then, the
resulting semilocal term in momentum space depends only on
$k_i$. This agrees with the AdS calculations: $\beta^{(i)}$ terms
are relevant whenever $\sigma\geq d/2+2\nu_i$. 

Next, let us consider the special situations. Logarithms
can arise from either the subdivergencies or the overall
divergence. 
\begin{enumerate}
\item If $\sum_i \omega_i = \sum_i \Delta_i = 2d + 2n$ there is an overall
divergence of degree $2n$. This leads to a simple logarithm at order
$\eps^0$ when there are no subdivergencies, and to higher powers of
logarithms if there are subdivergencies.
\item If $\omega_i - d=2n$, with $n\geq 0$ integer, the regularized
function contains a term with a logarithm at order $\eps^0$.
\item $\omega_i=0$ is the extremal case. We discuss it below.
\item If $\omega_i+\omega_j= 2 \Delta_k = d + 2n~(k\not \! = i,j)$
and $w_k\geq d$, the delta functions appearing in the regularization of
$x_{ij}^{-\omega_k}$  multiply $x_{ik}^{-2\Delta_k}$, which has degree
of divergence $2n$. Therefore, the completely regularized function
contains logarithms at orders $\eps^{2m-\omega+d}$ with
$m=0,1,\ldots,[\frac{1}{2}(\omega_k-d)]$. 
\end{enumerate}
These different possibilities perfectly agree, after the global rescaling, with
the respective ones analyzed in AdS. (We have used the same numbers
for each possibility in both descriptions.) 
Using the decomposition
\be
\Box \frac{\ln (x/\eps)}{x^2} =  \Box \frac{\ln (x M)}{x^2} + 4 \pi^2
\ln (M \eps) \delta(x) \, ,
\ee
we find the same powers of logarithms, with the same factors
$k_i^{2\nu_i}$. This equation is the CFT counterpart of the splitting
(\ref{splitlog}) in AdS. $\Box \frac{\ln (x M)}{x^2}$ should be
understood as a renormalized (sub)expression, with $M$ the
renormalization scale. 

The extremal case occurs when one $\omega_i$ vanishes. Then the
unregularized three-point function reduces to a product of
unregularized two-point functions. Suppose $\omega_3=0$. Then, 
\be
\la \O_{\Delta_1}(x_1) \O_{\Delta_2}(x_2) \O_{\Delta_3}(x_3) \ra = N
\frac{1}{x_{13}^{2\Delta_1}} \frac{1}{x_{23}^{2\Delta_2}} \, .  
\ee
This agrees with the structure of the nonlocal part of the AdS
results. Since bringing $x_1$ and $x_2$ together does not
lead to further divergencies, we only need to regularize
$1/x_{13}^{2\Delta_1}$ and $1/x_{23}^{2\Delta_2}$ independently.
From the AdS calculation we know that the regularization of each of
these functions coincides with the regularization of the two point
functions $\la \O_{\Delta_1} \O_{\Delta_1} \ra$ and $\la \O_{\Delta_2}
\O_{\Delta_2} \ra$, respectively. This does not occur in general for
subexpressions of nonextremal functions.

To end this section, let us consider briefly the renormalization of
the three-point functions. We do not need this for Randall-Sundrum but
it is relevant for the (renormalized) AdS/CFT correspondence. It is
clear that the overall divergence can be cancelled by a local
counterterm trilinear in the fields $\varphi$, similar to
(\ref{quadcounterterms}). 
The semilocal divergent
terms, on the other hand, cannot be cancelled by local counterterms
made out of fields $\varphi$ only. What we
need are counterterms that couple two fields $\varphi$ to operators
of the CFT. To see this, observe that the singular behaviour of the
three-point function 
when, say, $x_1\sim x_2$ and $x_3$ is kept appart, is given by the
terms with operators of dimension $\Delta_3$ in the OPE of 
$\O_{\Delta_1}(x_1)$ and $\O_{\Delta_2}(x_2)$:
\be
\O_{\Delta_1}(x_1) \O_{\Delta_2}(x_2) \sim
\ldots + \left[\frac{1}{x_{12}^{\omega_3}}\right]_\eps
O^\prime_{\Delta_3}(x_1) + \ldots \, ,
\ee
where $[\ldots]_\eps$ indicates regularization and $\O^\prime_{\Delta_3}$
denotes possible operators of dimension
$\Delta_3$, which can be different from the operators dual to
$\varphi_{\Delta_3}$ (typically they are double trace
operators). Agreement with AdS calculations 
implies that the regularization of $(1/x)^\omega$ is not universal,
\ie, it depends not only on $\omega$ but also on the particular
operators that appear in the 
OPE. For extremal correlators, nevertheless, we have seen that the
regularization of each factor is identical to the one in the two-point
functions. Since the relevant term in the OPE of both extremal three-point
functions (for the two lowest dimensional operators) and two-point
functions is the most singular one, this suggests that the regularization
of the most singular term of different OPEs (for fixed $\omega$) is
universal. The counterterms that cancel the semilocal
divergencies are, in a minimal substraction scheme,
\be
S^{\CFT\, {\rm ct}} = - \sum_{i,j} \int \dif^d x \dif^d y \varphi_i(x) \,
\left(\sum_{k} c_{ij}^k \O^\prime_k(x) {\rm div}
\left[\frac{1}{|x-y|^{\Delta_i+\Delta_j-\Delta_k}}\right]_\eps 
\right)\varphi_{j}(y)  + \ldots \, ,
\label{furthercounterterms}
\ee
where $c_{ij}^k$ are the Wilson coefficients in the OPE and div
denotes the part that diverges when $\eps\rightarrow 0$. Since the
latter is local, $S^{\CFT\, {\rm ct}}$ is a local functional. 
(\ref{furthercounterterms}) generalizes (\ref{quadcounterterms}).

%%%%%%%%%%%%%%%%%%%%%%%%%%%%%%%%%%%%%%%%%%%%%%%%%%

\section{Holographic description of RS1}
\label{RS1}

The so-called TeV brane of the RS1 scenario acts as a boundary for the
AdS geometry at large $z_0$. By the UV/IR correspondence, this must
correspond to some modification of the CFT such that
conformal invariance is broken in the IR. The scale below which the
deviation from conformal symmetry is significant is given by the
inverse of the position of the TeV brane, $z_0=\rho$.
The holographic dual of RS1 has been
studied recently in two interesting papers. In the
first one~\cite{Arkani-Hamed:2000ds}, N. Arkani-Hamed, M. Porrati and
L. Randall proposed some 
general features of the holographic theory and use them to explain
several phenomenological aspects of the model. In the second
one~\cite{Rattazzi:2001hs}, R. Rattazzi and A. Zaffaroni stablished
important details of the duality 
using the AdS/CFT correspondence and discussed some phenomenological
issues, like flavour symmetries. In this section we simply study
correlations functions of the induced fields in the presence of the
TeV brane.

The first question Rattazzi and Zaffaroni addressed was whether
conformal invariance is broken explicitly or spontaneously. An
explicit breaking by a relevant deformation would affect, to a certain
extent, the UV behaviour of correlators. In particular, the trace of
the stress-energy momentum would be modified. Rattazzi and Zaffaroni
showed that in the original model without stabilization the trace is
unchanged, so that the breaking is spontaneous. This agrees with the
discussion in~\cite{Balasubramanian:1999sn}, according to which a
change in the boundary 
conditions at large $z_0$ corresponds to a different vacuum in the
CFT. Moreover, Rattazzi and Zaffaroni found the corresponding 
Goldstone pole (associated to the radion, which is a modulus) using
both the effective Lagrangian of the radion and the 
AdS/CFT rules (what we called the D approach). So, conformal symmetry
is nonlinearly realized in the original RS1 (and further broken by the
UV cutoff). It was also checked in~\cite{Rattazzi:2001hs} that the
deviation from conformality of the nonlocal part of ($d$-dimensional)
massless two-point functions is exponentially suppressed for distances
$x \ll \rho$. This indicates that the operator adquiring a vev has
formally infinite conformal dimension. The Golberger-Wise
stabilization mechanism, on the other hand, introduces an explicit
breaking that in the holographic picture corresponds to a deformation
by an almost marginal deformation. This explicit breaking gives a small
mass to the radion. In order to generate a hierarchy, it is important
that the deformation be almost marginal. In the following we consider
the model without stabilizing scalars and refer
to~\cite{Arkani-Hamed:2000ds,Rattazzi:2001hs} 
for the interesting holographic interpretation of the Golberger-Wise
mechanism.

In RS1, the TeV brane does not alter the background metric in the
region between the two branes. But it does change the large $z_0$
boundary conditions of the fields propagating in this
background: since $\rho$ is an orbifold fixed point, fields must obey a
Neumann boundary condition at that point. Then, the propagator with
Dirichlet conditions at $\eps$ picks up a contribution from
$I_\nu$ Bessel functions: 
\be
\prop(z_0,\vec{k}) =  \frac{h(z_0)}{h(\eps)} \, ,
\ee
with
\bea
&& h(z_0)= z_0^{\frac{d}{2}} \left\{
\left[2 k\rho I_{\nu-1}(k\rho) + (d-2\nu) I_\nu(k\rho)\right] \, K_\nu(k
z_0) \right. \nn
&& \mbox{} + \left. \left[ 2 k\rho K_{\nu-1}(k\rho) + (2\nu-d) K_\nu(k\rho)
\right] \, I_\nu(k z_0) \right\}  \, .
\eea
Henceforth we take $L=1$.
The 1PI two-point function is obtained by inserting this propagator in
the quadratic action, which again reduces to the surface term:
\be
\la \varphi(\vec{k})\varphi(\vec{k})\ra^{\rm 1PI} = -\eps^{1-d}
\lim_{z_0\rightarrow \eps} \d_{z_0} \prop(z_0,\vec{k}) \, . \label{mod2pt}
\ee
For both points on the Planck brane and no boundary masses, its Minkowski
version is the inverse of the (Neunman) propagator calculated
in~\cite{Grinstein:2001ny,Gherghetta:2000kr}, as expected. The
explicit expression can be found in these
references. The Kaluza-Klein mass spectrum is now discrete
and can be obtained from the poles of this
propagator~\cite{Grinstein:2001ny}. Only 
massless bulk fields induce a zero mode on the brane. Negative
values of the squared bulk mass induce a single tachyon on the brane.
The couplings of the Kaluza-Klein modes at the brane
are given by the residues of the corresponding
poles~\cite{Arkani-Hamed:2000ds}. 

The expansion of (\ref{mod2pt}) around $\eps=0$ is
more intrincate than in the RS2 model. Instead of looking for a
generic expression, we have used Mathematica to calculate the
divergent (as $\eps\rightarrow 0$) local
terms and the first nonlocal term for several values of $\nu$. In all
cases we have found exactly the same divergent local terms as in the RS2
model. Moreover, the first nonlocal term is only
modified at scales $k>1/\rho$ by exponentially small corrections. For
integer $\nu$, the 
logarithm is the same as in the conformal case, but there are small
corrections to the local part at order $\eps^{2\nu-d}$. This is
related to the fact that the conformal anomaly is not altered by the
TeV brane. We have also checked numerically that the complete
two-point function is virtually independent of $\rho$ in the region $k
\gg 1/\rho$ as long as $\rho-\eps$ is not very small. 

The fact that divergent terms (after rescaling) are not affected by
the second brane is actually a particular case of a more general
principle: The divergent local part of the effective action in an 
asymptotically AdS space does not depend on perturbations at large
$z_0$~\cite{deHaro:2000wj}. It also agrees with the interpretation of
RS1 as a CFT in a nonconformal vacuum: the overall UV divergencies
are insensitive to the spontanous breaking, which only affects the IR
behaviour.

One could also compute three-point functions as in the RS2 model. The
local terms should be the same as in RS2. The semilocal terms, on the
other hand, will contain some dependence on $\rho$.

%%%%%%%%%%%%%%%%%%%%%%%%%%%%%%%%%%%%%%%%%%%%%%%%%%%

\section{Localized fields}
\label{localized}

In many phenomenologically interesting models, all or part of the
Standard model fields are constrained to live on either the Planck or
the TeV brane. This is the case of the original Randall-Sundrum
proposals, in which only gravity propagates in the bulk.
So it is important to incorporate these localized fields in the
effective descriptions\footnote{We use the term ``localized'' to refer
to fields that do not propagate in the extra dimension. They should
not be confused with Kaluza-Klein modes with bounded wave functions,
such as the graviton zero mode.}.  

The simplest possibility is having one field $\psi$ localized
on the Planck brane in either a RS1 or RS2 model. In general, $\psi$
will have some selfcouplings (including quadratic terms) and couplings
to the field $\phi$ restricted to the Planck brane. The effective
action of the whole system 
is simply the one obtained in the previous sections plus the $d$
dimensional action involving $\psi$ and its couplings to
$\varphi$. For example, if the $d+1$ dimensional action contains the
term
\be
S \supset \int_\eps \dif^{d+1} z \, \sqrt{g(z)} \psi(\vec{z})
\psi(\vec{z}) \phi(z) 
\delta(z_0-\eps) \, ,
\ee
the effective action will contain a local coupling $\int \dif^d z
\, \sqrt{g(\eps)} \psi(\vec{z}) \psi(\vec{z}) \varphi(\vec{z})$. In
particular, the 
exchange of $\varphi$ (with the propagator obtained above) will
contribute to the process $\psi\psi\rightarrow \psi\psi$. It is also
possible that $\phi$ itself has interactions localized on
the Planck brane. Again, these interactions should be simply added to
the effective action. One example is the case of boundary mass terms
discussed above. What about the holographic description? Since the
CFT only represents the bulk degrees of freedom, the
fields localized on the Planck brane cannot be part of the
CFT. Rather, they couple to it indirectly through $\varphi$.

The situation for fields localized at some $z_0>\eps$ is more
involved. We shall only find the effective theory in one
particular case and make a few qualitative comments about the
holographic interpretation. Consider in RS2 a coupling to an external
source located at $z_0=\rho$,
\be
S \supset - \int_\eps \dif^{d+1} z  \sqrt{g} j \phi(z) \delta(z_0-\rho)
\, . 
\ee
This is relevant to Lykken-Randall scenarios, in which the Standard
Model fields live on a probe three-brane of
infinitesimal tension that does not alter the AdS background. 
At the quadratic level, the bulk equation of motion in momentum space
is 
\be
z_0^{d+1} \d_0 \left( z_0^{-d+1} \d_0 \phi(z_0,\vec{k})\right) - (k^2
z_0^2 + m^2) \phi(z_0,\vec{k}) = j(\vec{k}) \delta(z_0-\rho) \, .
\ee
The  general solution to this equation can be written as 
\be
\phi(z_0,\vec{k}) = \phi^0(z_0,\vec{k}) + \sqrt{g(\rho)}
j(\vec{k}) \Delta_N(\rho;z_0,\vec{k}) \,
\ee
where $\phi^0$ is a solution to the homogenous equation and the
Neumann propagator $\Delta_N$ was defined above. The boundary
conditions $\phi(\eps,\vec{k})=\varphi(\vec{k})$,
$\lim_{z_0\rightarrow \infty} \phi(z_0,\vec{k})=0$ fix
\be
\phi^0(z_0,\vec{k}) = \left( \varphi(\vec{k}) - \sqrt{g(\rho)}
j(\vec{k}) \Delta_N(\rho;\eps,\vec{k}) \right) \, \prop(z_0,\vec{k})
\, . 
\ee
The action evaluated on this solution reduces to the surface term
\bea
S^\eff = \int \dif^{d} x  \dif^{d} y \, 
\left[\frac{1}{2} \varphi(\vec{x}) \left(- \eps^{-d+1} \frac{\d}{\d
\eps} \prop(\vec{x};\eps,\vec{y}) \right) \varphi(\vec{y}) + 
\varphi(\vec{x}) \rho^{-d-1}\prop(\vec{x};\rho,\vec{y}) j(\vec{y}) \right. \nn
\left. \mbox{} + \frac{1}{2} j(\vec{x}) \rho^{-2(d+1)}
\left( \int \dif^{d} z \Delta_N(\rho,\vec{x};\eps,\vec{z})
\prop(\vec{z};\rho,\vec{y}) - \Delta_N(\rho,\vec{x};\rho,\vec{y})
\right) j(\vec{y}) \right] \, . \label{jSeff}
\eea
We have used the identity (\ref{glue}). We see that the kinetic term
is not affected by the presence of the source, but there are
additional terms proportional to $\varphi j$ and $j^2$. From the
equations of motion of this action we find the on-shell brane field
\be
\varphi(\vec{k}) = \sqrt{g(\rho)} j(\vec{k})
\Delta_N(\rho;\eps,\vec{k}) \, , 
\ee
which agrees with the one obtained from the original $(d+1)$-dimensional
action. On the other hand, in the complete description one can think
of a field $\phi$ being generated by the source $j$,
propagating in $d+1$ dimensions and being eventually absorbed by $j$ at
$z_0=\rho$. The amplitud for this process is $g(\rho)
\Delta_N(\rho,\rho)$. This information is also contained in the
effective 
action:
\bea
\lefteqn{\left[ \frac{\delta}{\delta j(\vec{x_1})} \frac{\delta}{\delta
j(\vec{x_2})} \int \D \varphi e^{-S^\eff} \right]_{j=0}} && \nn 
&& = g(\rho) \int \dif^d y \, \dif^d z \,
\prop(\vec{x_1};\rho,\vec{y}) \la \varphi(\vec{y}) \varphi(\vec{z}) \ra_{j=0}
\prop(\vec{z};\rho,\vec{x_2})  \nn
&& \mbox{} - g(\rho) \left( \int \dif^{d} y
\Delta_N(\rho,\vec{x_1};\eps,\vec{y}) \prop(\vec{y};\rho,\vec{x_2}) -
\Delta_N(\rho,\vec{x_1};\rho,\vec{x_2}) \right) \nn 
&& =  g(\rho) \Delta_N(\rho,\vec{x_1};\rho,\vec{x_2}) \, .
\eea
In the last identity we have used (\ref{prop}) and (\ref{glue}).
In the holographic description, $j$ should couple to the CFT fields,
which are dual to the bulk degrees of freedom. However, $j$ changes
the equations of motion for $\phi$, which indicates that the conformal
symmetry of the holographic dual must be broken. In order to gain more
intuition about this point, it is useful to study the AdS theory from
the point of view of the probe brane. 

So far we have been discussing the effective theory for the field at
the Planck brane, which is obtained by putting $\phi(z)$ on shell for
$z_0>\eps$ and keeping $\varphi(\vec{z})=\phi(\eps,\vec{z})$ off
shell. Alternatively, we can 
find an effective theory for the field at the $d$-dimensional subspace
$z_0=\rho$. We
simply have to use in the action the equations of 
motion of $\phi(z)$ for $z_0\not \! = \rho$, with boundary
conditions
\bea
&& [\d_0 \phi(z)]_{z_0=\eps} = 0 \, , \nn
&& \lim_{z_0\rightarrow \infty} \phi(z) =0 \, , \nn
&& \lim_{z_0\rightarrow \rho} \phi(z) = \varphi_\rho(\vec{z}) \, .
\eea
These boundary conditions imply a discontinuity of $\d_0 \phi(z)$ at
$z_0=\rho$. We define the ``bulk-to-probe-brane'' propagators
$\prop^\rho_<$ and $\prop^\rho_>$ as solutions to the equations of
motion with boundary conditions (in momentum space)
\bea
&& [\d_0 \prop^\rho_<(z_0,\vec{k})]_{z_0=\eps} = 0 \, , \nn
&& \lim_{z_0\rightarrow \infty} \prop^\rho_>(z_0,\vec{k}) = 0 \, ,
\nn 
&& \prop^\rho_<(\rho,\vec{k}) = \prop^\rho_>(\rho,\vec{k}) = 1 \, .
\eea
Then, the on-shell field is $\prop^\rho_< \varphi_\rho$ ($\prop^\rho_>
\varphi_\rho$) for $z_0<\rho$ ($z_0>\rho$).
$\prop^\rho_>$ is given by (\ref{Dprop})
substituting $\eps$ by $\rho$, while $\prop^\rho_<$ is a combination of
the modified Bessel functions $K_\nu$ and $I_\nu$ that we do not write
explicitly. 
Inserting the on-shell $\phi$ into the action we find the
effective action
\bea
S_\rho^\eff [\varphi_\rho] & = & \frac{1}{2} \int_\eps \dif^{d+1} z \, \sqrt{g(z)} \d^\mu
(\phi(z) \d_\mu \phi(z)) \nn
& = & \frac{1}{2} \rho^{-d+1} \int \dif^{d} z \,
\phi(\rho) \left(\lim_{z_0\rightarrow \rho^-} \d_0 \phi(z) -
\lim_{z_0\rightarrow 
\rho^+} \d_0 \phi(z)\right) \nn
& = & \frac{1}{2} \rho^{-d+1} \int \dif^{d} x \,
\dif^{d} y \, \varphi(\vec{x}) 
\left[\d_0 \prop^\rho_<(\vec{x};z_0,\vec{y})- \d_0
\prop^\rho_>(\vec{x};z_0,\vec{y})\right]_{z_0-\rho} \varphi(\vec{y}) . \label{Srho}
\eea
The explicit result for the 1PI two-point function of the field
$\varphi_\rho$ in momentum space reads 
\bea
\lefteqn{\la \varphi_\rho(\vec{k}) \varphi_\rho(-\vec{k})\ra^{1PI}} &&
\nn
&& = \frac{\rho^{-d} \left[(\nu-\frac{d}{2}) K_\nu(k\eps) + k\eps
K_{\nu-1}(k\eps)\right]} 
{K_\nu(k\rho)\left[\left( (\frac{d}{2}-\nu) I_\nu(k\eps) + k\eps
I_{\nu-1}(k\eps)\right) K_\nu(k\rho) + \left((\nu-\frac{d}{2})
K_\nu(k\eps) + k\eps K_{\nu-1}(k\eps)\right) I_\nu(k\rho)\right]} \,
. \nn
\mbox{}
\label{rhotwopoint}
\eea
Now it is straightforward to add sources on the probe brane: they
couple linearly to $\varphi_\rho$, just as sources on the Planck brane
couple to $\varphi$. The inverse of (\ref{rhotwopoint}) gives the
propagator for the field $\varphi_\rho$, which is identical to
$\Delta_N(\rho;\rho,\vec{k})$ (and agrees
with~\cite{Giddings:2000mu} for massless fields). After Wick rotation
we find one pole at  
zero momentum for massless fields and one pole at spacelike
momentum for negative squared mass. For positive mass the
propagator contains no poles. On the other hand, the imaginary part is
positive definite and gives the amplitud at $z_0=\rho$ of the
Kaluza-Klein modes. This amplitud is oscillatory and vanishes at
discrete values of the Kaluza-Klein mass. Once more, we can expand the
two-point function around $\eps=0$. We find that the leading terms are
completely different from the ones in the expansion of the two-point
function on the Planck brane. These terms give the correct leading
behaviour of the propagator except in the neighbourhood of the pole.
The pole is a nonperturbative effect in $\eps$.

The effective theory $S_\rho^\eff$ is
useful to study physics on a probe brane at $z_0=\rho$. Its
holographic dual is obtained via the ``inner AdS/CFT correpondence''
proposed in~\cite{Balasubramanian:1999jd} (see
also~\cite{Csaki:2001cx} for an application to 
quasilocalized gravity scenarios). The idea is to decompose the
calculation in (\ref{Srho}) into two parts. First, we evaluate the
action with on-shell $\phi(z)$ only for $z_0<\rho$. The resulting
action depends on 
$\phi(z)$ at $z_0\geq \rho$: 
\be
S_\rho = \tilde{S}^\eff_\rho[\varphi_\rho] + \frac{1}{2} \int_\rho
\dif^{d+1} z \, \sqrt{g(z)}  \left(\d^\mu \phi \d_\mu \phi + M^2
\phi^2 \right) \, , \label{intstep}
\ee
where 
\be
\tilde{S}^\eff_\rho[\varphi_\rho] = \frac{1}{2}
\rho^{-d+1} \int \dif^{d} x \, 
\dif^{d} y \, \varphi_\rho(\vec{x}) \left[\d_0
\prop^\rho_<(\vec{x};z_0,\vec{y})\right]_{z_0=\rho}
\varphi_\rho(\vec{y}) 
\label{complicado}
\ee
is an effective action describing the effect of the bulk degrees of
freedom between $\eps$ and $\rho$. The second term in (\ref{intstep})
is simply the action of an RS2 model with Planck brane at
$\rho$. It is therefore dual to the same CFT with regulator $\rho$,
perturbed by operators coupled to $\varphi_\rho$. The contribution
$\tilde{S}^\eff_\rho$ arises from integrating out the degrees of
freedom of the CFT heavier than $1/\rho$. As we can see in
(\ref{complicado}), the (Wilsonian) renormalization group flow of
the theory from $1/\eps$ to $1/\rho$ is not the simple rescaling one
would expect in a conformal theory. A more complicated structure
arises due to the Dirichlet boundary condition at $z_0=\rho$, which
substitutes the condition of regularity at
infinity\footnote{In~\cite{Balasubramanian:1999jd} a conformal flow
given by $\prop_\nu(\rho)$ is found. The 
reason is that no condition is imposed at the boundary, and
the derivative of the on-shell field is implicitly assumed to be
continuous at $z_0=\rho$. We observe that if the Planck brane is
removed, one should impose regularity of the field as it approaches
the AdS boundary. This implies a discontinuous derivative at
$z_0=\rho$.}. A change of boundary condition inside AdS
corresponds to a different vev for the dual operator, \ie, to a change
of vacuum in the CFT~\cite{Balasubramanian:1999sn}. The situation we
are describing is then  
consistent with the interpretation in~\cite{Arkani-Hamed:2000ds} of
the holographic dual of Lykken-Randall (for the Standar type IIB
correspondence) as arising from Higgsing the original CFT: $U(N)
\rightarrow U(N-1)\times U(1)$. If this is correct, the RG flow 
leading to (\ref{complicado}) is the flow from $1/\eps$ to
$1/\rho$ of a CFT with a Coulomb branch deformation (in the
approximation where the metric is fixed). We have mentioned that the
first terms in the small $\eps$ expansion of the $\varphi_\rho$ two-point
function differ from the ones of the $\varphi$ two-point
function. This indicates that the regulator structure is mixed in a
nontrivial way by the RG flow. Note that once we reach $\rho$,
the effective action dual to $S_\rho$ is given by the nonlocal action
$\tilde{S}_\rho^\eff$ coupled to a CFT in the conformal phase with
operator deformations. If the brane at $z_0=\rho$ is the TeV brane of
RS1, the space terminates at $\rho$ and the whole effective
theory at that brane is described by $\tilde{S}_\rho^\eff$ alone.

In the holographic interpretation we have proposed, using
(\ref{intstep}) as guiding principle, the correct propagator is
reproduced when the source couples directly to $\varphi_\rho$, as long
as the RG flow gives the bilinear term (\ref{complicado}) (in a low
energy approximation). On the other hand, it has been proposed
in~\cite{Balasubramanian:1999sn} (see also the earlier discussion
in~\cite{Giddings:2000ay}) that the effective theory on the brane can
contain an additional coupling of the source to a CFT operator (but no
bililear term in 
$\varphi_\rho$): 
\be
\int \rho^{2\Delta-d} j \O_\Delta \, \label{jo} \, .
\ee
Integrating out the remaining CFT modes one finds, schematically, 
\be
\varphi_\rho \la \O \O \ra_\rho \varphi_\rho + \rho^{2\Delta-d} j \la
\O \O \ra_\rho \varphi_\rho + \rho^{4\Delta-2d} j \la \O \O \ra_\rho j
\, , 
\label{alli}
\ee
which is analogous to (\ref{jSeff}) and gives the right qualitative
low-energy behaviour of the propagator. However, we cannot reproduce
in this way the exact propagator since (\ref{jo}) and (\ref{alli})
contain no information about $\eps$. Furthermore, we have not found
a definite AdS counterpart of the term (\ref{jo}). So, it seems
difficult to reconciliate this interpretation with the regularized
AdS/CFT correspondence.

\section{Discussion}
\label{discussion}

We have calculated generic two-point and three-point correlation
functions of induced fields in Randall-Sundrum models (mainly RS2),
and shown explicitly that they can be obtained holographically from a
{\em regularized} CFT. Local and semilocal terms for the brane fields
arise from the regularization of the theory. These terms are more
important at low energies than the nonlocal ones arising from the
unregulated CFT.
Alternatively, Randall-Sundrum models can be
holographically described as a {\em renormalized} CFT coupled to a
local theory for 
the induced fields: $S^{\CFT}[A]+ S^\prime[A,\varphi]$.
$S^\prime[A,\varphi]$ contains the couplings $\varphi \O$ and more
complicated couplings that are related to the IR counterterms of the AdS
theory. In particular it contains kinetic terms for
$\varphi$, which can then be naturally treated as a dynamical
field. In fact, this is what one should do, since in the AdS theory the
induced fields are dynamical and not just external sources. In other
words, in the complete theory one has to include both sides
of~(\ref{regAdSCFTrelation}) inside a path integral over
$\varphi$. Although in this paper we  
have followed the ``regularization'' point of view, the
``renormalization'' approach is more adequate in some discussions,
particularly when the 
holographic theory is taken as the starting
point~\cite{Frampton:1999yb,Frampton:2001kk} and AdS 
is just a calculational tool. Indeed, from the point of view of the
quantum field theory it is more natural to say that the world is
described by a particular renormalized theory, rather than by a theory
regularized in a very particular way. From this viewpoint, the Planck
brane should not be thought of as a regulator, but rather as an
element one has to add to the AdS description to take into account 
the effect of $S^\prime$. One can then do model building modifying
$S^\prime$, which will correspond to changing the Planck brane by a more
complicated object.

For simplicity we have studied only scalar fields. Most of the results
generalize to fields with higher spin. The effective theories for the
graviton and gauge fields will have gauge symmetries, since a subgroup
of the $d+1$-dimensional diffeomorphism and gauge invariance acts on
the field induced on the brane. In the usual AdS/CFT
correspondence, gauge symmetries in the bulk induce global symmetries
in the boundary theory. In the holographic Randall-Sundrum theory,
these global symmetries are gauged by the induced fields on the
brane. An essential point here is that the regularization (or 
renormalization) of the CFT respects the Ward identities of these
symmetries. The Planck brane provides such a regularization (respecting
gauge invariance and supersymmetry, which is a hard task from a
field-theoretical approach). On the other hand, knowing that
the effective theory has a nontrivial symmetry is very useful because
the interacting terms are related to the quadratic ones, which are
much simpler to calculate.

It is important to examine the approximations and assumptions we have
used in this paper. First, we have been mainly interested in a low-energy
expansion. For momenta smaller than the inverse AdS radius, the effect
of the regularization is local to a good approximation, and can be
interpreted as the addition of a local action $S^\prime$ to the
renormalized CFT. Higher order effects can be reproduced by a
nonlocal $S^\prime$, but this looks like a rather artificial
construction from the field theory side. We observe that this
approximation is related to the way in which the Planck brane modifies
the standard AdS/CFT correspondence. So, considering string theory
instead of supergravity (or small coupling in the field theory) does
not help in going beyond the low-energy approximation.

Second, we have ignored throughout the paper the
back-reaction of the scalar fields on 
the metric. If the back-reaction is small, it will only affect the
fluctuations of the metric (the graviton), which can be perturbatively
studied with the same formalism. However, for some ranges of parameters the
back-reaction changes dramatically the AdS background. This effect can
be important for the scalars stabilizing the two-brane
models~\cite{DeWolfe:2000cp,DeWolfe:2000cp}. Moreover, it has been shown
in~\cite{DeWolfe:2000xi} that generic solutions preserving
four dimensional Poincar\'e have a positive-definite potential in the
equivalent quantum-mechanical problem, and therefore there are no
tachyons in the effective theory on the boundary. Presumably this
result applies also to the effective theory at finite $\eps$,
showing that the instability found in~\cite{Ghoroku:2001pi} and here
for negative squared mass is just
an artifact of the approximation of ``inert'' scalars. The divergencies
of the coupled scalar-gravity system have been calculated
in~\cite{deHaro:2000wj} 
for scalars with $\Delta\leq d$ (corresponding to relevant
or marginal deformations), while for $\Delta>d$ one 
finds uncontrollable divergencies and the formalism of~\cite{deHaro:2000wj}
cannot be applied. Correlation functions of active scalars
were first studied in~\cite{DeWolfe:2000cp}, and
it was found that the usual AdS/CFT prescription could not be used
when the scalars are not decoupled. Subsequently, a method was
developed in~\cite{Arutyunov:2000rq} to decouple the scalar
fluctuations from the graviton ones. This allowed the authors of this
paper to compute the two-point functions of the decoupled scalar in
the usual way (see also~\cite{Bianchi:2001sm}). It would be interesting
to apply this method to brane world models. 

Third, in our CFT calculations we have always treated $\varphi$
as an external field and computed $\varphi$ correlators ignoring the
effect of virtual $\varphi$ interacting with the CFT fields. Since the
CFT fields are dual to the bulk degrees of freedom, including such
quantum corrections should correspond to a dynamical modification
of the AdS background. One might hope 
that the full quantum theory $S^\CFT+S^\prime$  is related somehow to
the exact coupled gravity-scalar theory in the presence of the Planck
brane, but we have not found convincing arguments supporting this
idea. At any rate it is clear that if one starts with the
$d$-dimensional theory $S^\CFT+S^\prime$, the effect of $S^\prime$ in
quantum corrections is important. This issue has been briefly discussed
in~\cite{Arkani-Hamed:2000ds}. There it has been pointed out that, in
order to agree with the AdS picture, $S^\prime$ must be such that the
theory remains conformal in the IR. To protect the scalars of the CFT
from getting a 
mass through radiative corrections, one needs to impose some symmetry
on $S^\prime$, such as supersymmetry, that forbids strongly relevant
operators (nearly marginal operators can give rise to a
Goldberger-Wise stabilization mechanism). We finish these comments 
observing that the AdS problems with $\Delta>d$ are
related to the fact that the perturbed theory
is in this case nonrenormalizable and requires an infinite set of
counterterms. 

Our analysis of the holographic duals of RS1 and Lykken-Randall models
has been less quantitative and we have left many open questions. A more
complete study requires a better understanding of the inner AdS/CFT
correspondence and its relation with the RG equations satisfied by the
correlation functions.

%%%%%%%%%%%%%%%%%%%%%%%%%%%%%%%%%%%%%%

\section*{Acknowledgments}

I would like to thank A. Awad, J. Erdmenger, E. Katz, A. Montero,
L. Randall and M. Tachibana for valuable discussions. I acknowlege
financial support by MECD.

%%%%%%%%%%%%%%%%%%%%%%%%%%%%%%%%%%%%%%%%%%

\section*{Appendix}

\renewcommand{\theequation}{A.\arabic{equation}}

In this appendix we collect several formulae that are useful in
differential regularization. Then, we give an explicit example of the
regularization of a three-point function. 

Differential identities in $d$ dimensions for $x\not \! = 0$:
\bea
\frac{1}{x^{\beta}} &=& \frac{1}{(2-\beta)(\beta-d)} \Box
\frac{1}{x^{\beta-2}} \, ,~\beta > d \, ,  \\ 
\frac{1}{x^d} &=& \frac{1}{(2-d)} \Box
\frac{\ln(xM)}{x^{d-2}} \, ,\\
\Box \frac{1}{x^{d-2}} &=& - \frac{4\pi^{d/2}}{\Gamma(\frac{d}{2}-1)} \,
. \label{delta}
\eea
These identites can be used iteratively to find the regularized value
of two-point functions that we have written in Section \ref{CFT}. For
the example below we also need the following identities (for $x\not \!
= 0$). We only write them in 4 dimensions: 
\bea
\frac{\ln(xM)}{x^4} &=& -\frac{1}{4} \Box
\frac{\ln^2(xM)+\ln(xM)}{x^2} \, ,  \\ 
\frac{\ln(xM)}{x^6} &=& -\frac{1}{64} \Box \Box
\frac{2\ln^2(xM)+5\ln(xM)}{x^2} \, .
\eea
To find the Fourier transforms of the regularized functions we only
need the ``integration by parts'' rule of differential regularization
and the following Fourier transforms:
\bea
\frac{1}{x^{2\alpha}} &\longrightarrow & \frac{2^{d-2\alpha} \pi^{d/2}
\Gamma(\frac{d}{2}-\alpha)}{\Gamma(\alpha)}\, k^{2\alpha-d} \, ,
\alpha<\frac{d}{2} \, , \\
\frac{\ln(xM)}{x^{d-2}} &\longrightarrow &
\frac{2\pi^{d/2}(2-d)}{\Gamma(\frac{d}{2})}
\frac{\ln(k/\bar{M})}{k^2} \, ,
\eea
with $\bar{M}=\frac{2M}{\gamma_E}e^{\psi(\frac{d}{2}-1)}$,
$\gamma_E=1.781\ldots$ and $\psi$ being the Euler constant and Euler
psi function, respectively.

Let us consider now the regularization of the three-point correlator
with $\Delta_1=\Delta_2=3$, $\Delta_3=4$, in dimension $d=4$. For
noncoincident points, 
\be
\la \O_3(x_1) \O_3(x_2) \O_4(x_3) \ra = N \frac{1}{x^4 y^4 (x-y)^2} \, ,
\ee
where $x$ and $y$ were defined in the text.
First, we have to regularize the divergencies at $x\sim 0$ and $y\sim
0$ for $x\not \! = y$. We find
\be
\left(-\frac{1}{2}\Box_x \frac{\ln x/\eps}{x^2} + b_0^{(1)} \delta(x)
\right) \left(-\frac{1}{2}\Box_y \frac{\ln y/\eps}{y^2} + b_0^{(2)} \delta(y)
\right)\frac{1}{(x-y)^2} \, .
\ee
Using the Leibniz rule and (\ref{delta}) we can
write this expression as a total derivative of a well-behaved
distribution plus divergent terms depending on just one variable
(times deltas):
\bea
&&\frac{1}{4} \Box_x \Box_y \frac{\ln(x/\eps)\ln(y/\eps)}{x^2y^2(x-y)^2}
- \pi^2 \Box_x \left[ \frac{\ln^2 (x/\eps)}{x^4} \delta(x-y)\right]  \nn
&& \mbox{} - \frac{b_0^{(2)}}{2} \Box_x \frac{\ln(x/\eps)}{x^4} \delta(y)  
- \frac{b_0^{(1)}}{2}
\Box_y \frac{\ln(y/\eps)}{y^4} \delta(x) + 2\pi^2 \frac{\ln
(x/\eps)}{x^6} \delta(x-y) \nn
&& \mbox{} + \left(2\pi^2 b_0^{(2)} \frac{\ln (x/\eps)}{x^2}
+b_0^{(1)} b_0^{(2)} \frac{1}{x^2} \right) \delta(x)\delta(y) 
\eea
Finally, we regularize the overall divergence using the differential
regularization identities above. The final regularized
function reads
\bea
&& \la \O_3(x_1) \O_3(x_2) \O_4(x_3) \ra = \frac{1}{4} \Box_x \Box_y
\frac{\ln(x/\eps)\ln(y/\eps)}{x^2y^2(x-y)^2} \nn
&& \mbox{} + \frac{\pi^2}{4} \Box_x (\d_x+\d_y)^2
\left[\frac{\ln^2(x/\eps)+\ln(x/\eps)}{x^2} \delta(x-y)\right] \nn
&& \mbox{} + \frac{b_0^{(2)}}{8} \Box_x^2
\frac{\ln^2(x/\eps)+\ln(x/\eps)}{x^2} \delta(y) + \frac{b_0^{(1)}}{8}
\Box_y^2 \frac{\ln^2(y/\eps)+\ln(y/\eps)}{y^2} \delta(x) \nn
&& \mbox{} - \frac{\pi^2}{36} (\d_x+\d_y)^4 \left[
\frac{2\ln^2(x/\eps)+5\ln(x/\eps)}{x^2} \delta(x-y) \right] \nn
&& \mbox{} + \left(a_0 \frac{1}{\eps^2} + a_1^{(1)} \Box_x + a_1^{(2)}
\Box_y) + a_1^{(3)} (\d_x+\d_y)^2 \right)
\left[\delta(x)\delta(y)\right] \, . 
\eea
Observe that $\d_x$, $\d_y$ and $\d_x+\d_y$ Fourier transform into
$-i k_1$, $-i k_2$ and $-i k_3= -i(k_1+k_2)$, respectively. We could
also have obtained a longer expression explicitly symmetric under $x_1
\leftrightarrow x_2$.

\vspace*{2cm}

%%%%%%%%%%%%%%%%%%%%%%%%%%%%%%%%%%%%%%

\end{document}